\documentclass[12pt]{article}

\usepackage[latin2]{inputenc}
\usepackage[usenames]{color}
\usepackage{latexsym}
\usepackage{epsf}
\usepackage{epsfig}
\usepackage{amssymb,amsmath}

\def\nn{\nonumber \\}

\def\cC{{\cal C}}
\def\cP{{\cal P}}
\def\cR{{\cal R}}
\def\cL{{\cal L}}
\def\cT{{\cal T}}

\def\cM{{\cal M}}
\def\cN{{\cal N}}
\def\Z2{{\mathbb Z}_2}

\def\pa{\partial}
\def\Ln{{\pounds_n}}

\def\ocT{{\overline{\cal T}}}

\def\beq{\begin{equation}}
\def\eeq{\end{equation}}

\def\bea{\begin{eqnarray}}
\def\eea{\end{eqnarray}}

\def\bi{\begin{itemize}}
\def\ei{\end{itemize}}

\def\be{\begin{enumerate}}
\def\ee{\end{enumerate}}

\def\bc{\begin{center}}
\def\ec{\end{center}}

\newlength{\dinwidth}
\newlength{\dinmargin}
\begin{document}

\thispagestyle{empty}
%{\color{red} version: \today }
\begin{flushright}
IFT--09--7\\
\end{flushright}

\vspace*{15mm}

\begin{center}
{\Large\bf 
Higher order dilaton gravity: 
\\
brane equations of motion in the covariant formulation
}
\vspace*{5mm}
\end{center}

\vspace*{5mm} \noindent
\vskip 0.5cm
\centerline{\bf
Dominika Konikowska\footnote[1]{Dominika.Konikowska@fuw.edu.pl}
and Marek Olechowski\footnote[2]{Marek.Olechowski@fuw.edu.pl}
}
\vskip 5mm
\centerline{\em Institute of Theoretical Physics,
University of Warsaw}
\centerline{\em ul.\ Ho\.za 69, PL--00--681 Warsaw, Poland}

\vskip 15mm

\centerline{\bf Abstract}
\vskip 3mm

Dilaton gravity with general brane localized interactions is investigated.  
Models with corrections up to arbitrary order in field derivatives are 
considered. Effective gravitational equations of motion at the brane are 
derived in the covariant approach. Dependence of such brane equations on 
the bulk quantities is discussed. It is shown that the number of the bulk 
independent brane equations of motion depends strongly on the symmetries 
assumed for the model and for the background. Examples with two and four 
derivatives of the fields are presented in more detail.

\newpage

%%%%%%%%%%%%%%%%%%%%%%%%%
\section{ Introduction }
%%%%%%%%%%%%%%%%%%%%%%%%%

Assuming that the space-time dimensionality is restricted 
to the number of currently known 4 dimensions, the Einstein 
tensor appearing in the equation of motion of the standard 
theory of gravity is the most general rank 2, divergence-free 
symmetric tensor depending on the metric and its first and 
second derivatives only, while being linear in the latter.
As for the presently testable, gravity-related phenomena,
all experiments seem to show no 
discrepancies between the General Relativity and 
observations\footnote{
However, there are claims that some alternative 
models could provide better explanation for the 
late-time cosmic acceleration
%accelerated expansion of the late Universe  
and/or for the phenomena attributed 
usually to the existence of the Cold Dark Matter.
(see e.g.\ \cite{alternatives} and references therein).
}. 
Nevertheless, a beyond 
the Standard Model theory is commonly pronounced necessary, 
with string theories amongst the most serious candidates. 
As compared to the standard Einstein theory of gravity, the 
effective Lagrangians obtained within the string theories framework 
contain corrections with higher powers of the Riemann tensor 
\cite{MeTs0GrSl}. Moreover, the $\alpha'$ expansion in string 
theories predicts corrections of the higher order in derivatives 
not only for the metric tensor, but also for other fields, such as 
the dilaton. In this work we investigate models with higher order 
corrections for both the metric and the dilaton, as well as mixed 
gravity-dilaton interactions.

The order of the corrections is restricted by the dimensionality 
of the space-time. Specifically, the $N$-th power of the Riemann 
tensor can be included into a dilaton gravity Lagrangian only if 
the number of space-time dimensions, $d$, is sufficient, i.e.\ for 
$d\ge 2N$. For example, the 2nd order contribution quadratic in 
the Riemann tensor, known as the Gauss-Bonnet (GB) term \cite{La}, 
can be taken into account in a dilaton gravity theory already in 
4 dimensions. On the other hand, in the classical, no-dilaton theories, 
the GB term is a full divergence in 4 dimensions. It becomes 
dynamically relevant in an at least 5-dimensional space-time only. 
For pure gravity the GB term was generalized to higher 
orders\footnote{
Explicit formulae for the 3rd and 4th orders expressions
are given in \cite{Mu} and \cite{Wh}, respectively.
} 
by Lovelock \cite{Lo}. 
In a previous paper by the present authors \cite{KoOl},
the Einstein-Lovelock theory of gravity was generalized by 
coupling it to the dilaton. The appropriate action and equations 
of motion were constructed up to arbitrary order in derivatives 
of both the metric tensor and the dilaton.

The main motivation for the dilaton gravity theory constructed 
in \cite{KoOl} comes from string theories, which 
are considered to be the most promising attempt to quantize 
gravity and unify it with all other known interactions. 
The gravitational sectors of the effective field theories derived 
from string theories coincide with the Einstein theory of gravity only  
in the lowest order. Beyond such approximation they involve   
higher powers of the Riemann tensor as well as 
interactions with other fields with more than two derivatives.  
The dilaton gravity model considered in the present paper is closely 
connected to string theories - its interactions with up to four 
derivatives are exactly as those present in the effective theory 
derived from string theories and restricted to the gravity and the
dilaton field (see e.g.\ \cite{Me}). Such correspondence has not been 
proven for interactions with six or more derivatives. However, there 
are considerable indications (see the discussion of the $O(d,d)$ symmetry in 
\cite{KoOl}) that our model may be a part of the effective string 
dilaton gravity action also at the level of more than four derivatives.

Dating back to the M-theory based work by Ho\v{r}ava and Witten 
\cite{HoWi}, the idea of localizing the Standard Model on a brane  
embedded in a higher-dimensional space-time \cite{Ak0RuSh} 
gained quite a lot of 
attention. Corrections to the gravity interactions at the brane 
due to the bulk fields were investigated by many authors \cite{review}. 
It seems interesting and important to consider what gravity would 
be induced at the brane, if the bulk action was given by a higher 
order dilaton gravity theory. 
Hence, the purpose of this work is to derive 
effective gravitational equations at the brane 
for the models of arbitrary higher order in derivatives, 
constructed in \cite{KoOl}. 
In order to keep full generality, the procedure will be carried out 
in the covariant approach. Starting from the $d$-dimensional ($d>4$) 
higher order dilaton gravity theory, equations of motion 
in the effective $(d-1)$-dimensional theory, i.e.\ for a co-dimension 
1 brane, will be derived.

For the standard lowest order gravity the effective equations of 
motion at the brane  were derived in the covariant approach in 
\cite{SaShMa}. That analysis was extended to the GB gravity in 
\cite{MaTo}. The effective equations at the brane for the lowest order 
dilaton gravity were derived in \cite{MaWa} (however, not in a fully 
covariant way). The covariant approach was employed in \cite{MeBa}
for cosmological applications of dilaton gravity.
Certain second order gravity models with branes, including either first 
or second order corrections for the scalar field, were analyzed for 
specific metrics in e.g.\ \cite{di1gra2} and \cite{di2gra2}, respectively. 
Although brane models for arbitrary order Einstein-Lovelock gravity 
were investigated in \cite{MeOl}, neither the dilaton field was included,
nor the covariant approach was adopted. No comprehensive results for 
theories which simultaneously take into account interactions of the 
higher order in the Riemann tensor and involve the dilaton have 
been presented so far.

The main goal of the present work is to 
obtain the effective brane equations of motion for arbitrary 
order dilaton gravity models using the covariant approach.
Due to the already pointed out relation of the higher order dilaton 
gravity model \cite{KoOl} to string theories, the covariant derivation 
of the effective brane equations in this setup will allow for studying
in a more systematic way the potentially observable effects of string 
theories e.g.\ on specific cosmological models.

The rest of this paper is organized as follows: 
In section \ref{section_higher_dim_EoM} we define the $d$-dimensional 
dilaton gravity models 
with corrections of the higher order in derivatives and with brane interactions
whose exact nature we choose not to specify. 
The bulk equations of motion for these models are derived 
and rewritten in terms of quantities either projected on the brane, 
or perpendicular to it. 
Section \ref{section_JC} is devoted to the analysis of the 
junction conditions at the brane. 
The effective equations of motion at the brane for a general 
case with arbitrary order corrections 
are obtained in section \ref{section_effectiveEoM}.
Construction of all possible brane equations is carefully discussed. 
The importance of the bulk $\Z2$ symmetry for the form and existence 
of the effective equations is analyzed. 
In section \ref{section_examples} explicit results for the 
lowest order dilaton gravity 
and for the theory with up to four derivatives are presented. 
Some of the results for the latter, because of their complexity, 
are moved to the appendix. 
Section \ref{section_conclusions} contains our conclusions.

%%%%%%%%%%%%%%%%%%%%%%%%%%%%%%%%%%%%%%%%%%%%%%%%%%%%%%%%%%%%%%%%
\section{Higher dimensional equations of motion}
\label{section_higher_dim_EoM}
%%%%%%%%%%%%%%%%%%%%%%%%%%%%%%%%%%%%%%%%%%%%%%%%%%%%%%%%%%%%%%%%

We consider the $d$-dimensional dilaton gravity theory described 
by the following Lagrangian:
\begin{equation}
\cL 
=
%  &=& 
e^{-\phi}
\left[
\sum_{N=1}^{N_{max}} \frac{\alpha_N}{2}\, 
\cT \left( \Big[ {\textstyle\frac12} \cR_{**}^{**}
    \oplus 2 (\nabla\nabla)_*^*\phi \oplus (-1)(\pa\phi)^2 \Big]^N \right) 
-V(\phi)  +  \cL_B \, \delta_B \right]
.
\label{L}
\end{equation}
This is the Lagrangian (in the notation explained below) of the 
higher order dilaton gravity constructed in \cite{KoOl} and 
generalized by including a bulk scalar potential $V(\phi)$ and 
general brane interactions given by $\cL_B$. The position of the 
brane\footnote{
The theory can include more branes,
but we choose to focus our considerations on the brane
for which we will be deriving the effective gravitational equations of motion. 
} 
is described by the Dirac delta type distribution $\delta_B$. 
As was discussed in the Introduction, the main motivation for 
considering this dilaton gravity model is its close relation to 
string theories. Investigation of its properties in a brane 
scenario is a step towards studying e.g.\ the potentially 
observable cosmological effects of string theories.

The above Lagrangian is written in a very compact form using 
the following notation introduced in \cite{KoOl}: $\cT$ is 
a generalization of the ordinary trace. Acting on an arbitrary 
rank $(m,m)$ tensor $M$ it returns  a number given by
\begin{equation}
\cT(M)
=
\delta_{\rho_1 \rho_2 \cdots \rho_m}^{\sigma_1 \sigma_2 \cdots \sigma_m}
M^{\rho_1 \rho_2 \cdots \rho_m}{}_{\sigma_1 \sigma_2 \cdots \sigma_m}\,,
\label{cT}
\end{equation}
where $\delta$ with $m$ pairs of indices is the generalized 
Kronecker delta
\begin{equation}
\delta_{\rho_1 \rho_2 \cdots \rho_m}^{\sigma_1 \sigma_2 \cdots \sigma_m}
=
\det\left[
\begin{tabular}{cccc}
$\delta_{\rho_1}^{\sigma_1}$ & $\delta_{\rho_2}^{\sigma_1}$ & $\cdots$
& $\delta_{\rho_m}^{\sigma_1}$\\
$\cdot$ & $\cdot$ & $\cdots$ & $\cdot$\\
$\cdot$ & $\cdot$ & $\cdots$ & $\cdot$\\
$\delta_{\rho_1}^{\sigma_m}$ & $\delta_{\rho_2}^{\sigma_m}$ & $\cdots$ 
& $\delta_{\rho_m}^{\sigma_m}$\\
\end{tabular}
\right] .
\label{delta}
\end{equation}
Asterisks are used as indices in the Lagrangian (\ref{L}) to 
indicate ranks of tensors (e.g.\ to distinguish the Riemann 
tensor from the Ricci tensor or the curvature scalar). 
Under the generalized trace $\cT$ there are powers of ``sums'' 
(denoted by $\oplus$)  - i.e.\ of linear combinations of different 
ranks tensors. 
This should be understood as a compact notation for the following 
operation: First, we perform the algebraic manipulations 
(sums and powers), treating all tensors as ordinary numbers. 
Then the generalized trace (\ref{cT}) of the obtained linear 
combination of tensors products should be understood as the 
corresponding linear combination of the generalized traces of 
tensors products. For example:
\begin{eqnarray} 
&& \hspace{-0.5cm}
\cT
\left( \Big[ {\textstyle\frac12} \cR_{**}^{**}
    \oplus 2 (\nabla\nabla)_*^*\phi \Big]^2 \right)
=
{\textstyle\frac14} \, \cT\big(\cR_{**}^{**}\cR_{**}^{**}\big)
+2 \, \cT\big(\cR_{**}^{**}(\nabla\nabla)_*^*\phi\big)
+4 \, \cT\big((\nabla\nabla)_*^*\phi(\nabla\nabla)_*^*\phi\big) 
\nonumber
\\[4pt]
&& =
{\textstyle\frac14}\,
\delta_{\rho_1\rho_2\rho_3\rho_4}^{\sigma_1\sigma_2\sigma_3\sigma_4}
\,\cR^{\rho_1\rho_2}{}_{\sigma_1\sigma_2} \,
\cR^{\rho_3\rho_3}{}_{\sigma_3\sigma_4}
+2\,
\delta_{\rho_1\rho_2\rho_3}^{\sigma_1\sigma_2\sigma_3}
\,\cR^{\rho_1\rho_2}{}_{\sigma_1\sigma_2} 
\left(\nabla^{\rho_3}\partial_{\sigma_3}\phi\right)
+4\,
\delta_{\rho_1\rho_2}^{\sigma_1\sigma_2}
\left(\nabla^{\rho_1}\partial_{\sigma_1}\phi\right) 
\left(\nabla^{\rho_2}\partial_{\sigma_2}\phi\right)
,
\nonumber
\end{eqnarray}
where, due to the properties of the Riemann tensor 
and the second covariant derivative of a scalar, we can 
unambiguously define: 
$\cR_{\mu\nu}^{\rho\sigma}\equiv{\cR^{\rho\sigma}}_{\mu\nu}$
and $(\nabla\nabla)_\mu^\nu\phi\equiv\nabla^\nu\partial_\mu\phi$.

With the above formulae it is easy to check that 
the number of higher order terms in the Lagrangian (\ref{L}) 
is not arbitrary. The term in (\ref{L}) with the generalized 
trace (\ref{cT}) of the highest rank tensor is proportional to 
\begin{equation}
\cT\left(\left(\cR_{**}^{**}\right)^{N_{\rm max}}\right)
=
\delta_{\rho_1\rho_2\cdots\rho_{(2N_{\rm max}-1)}\rho_{(2N_{\rm max})}}
^{\sigma_1\sigma_2\cdots\sigma_{(2N_{\rm max}-1)}\sigma_{(2N_{\rm max})}}
\cR^{\rho_1\rho_2}{}_{\sigma_1\sigma_2}\ldots
\cR^{\rho_{(2N_{\rm max}-1)}\rho_{(2N_{\rm max})}}
{}_{\sigma_{(2N_{\rm max}-1)}\sigma_{(2N_{\rm max})}}
\,.
\label{highest}
\end{equation}
It is obvious from the definition (\ref{delta}) that, 
because of the antisymmetry in all indices of one 
type (covariant and contravariant), the $(2N_{\rm max}, 2N_{\rm max})$ 
rank generalized Kronecker delta (\ref{delta}) in eq.\ (\ref{highest}) 
is non-zero only if the number of 
space-time dimensions $d$ is sufficiently large. 
Thus, there is an upper limit on the order of the interaction 
terms present in our Lagrangian:  $N_{\rm max}\le(d/2)$.

Using the results of \cite{KoOl}, 
bulk equations of motion can be derived from the Lagrangian 
(\ref{L}). They read 
\begin{eqnarray}
  g_{\mu\nu} V(\phi)
  - \sum_{N=1}^{N_{max}} \frac{\alpha_N}{2}\, {\ocT}_{\mu\nu} 
\left( \Big[ {\textstyle\frac12} \cR_{**}^{**}
\oplus 2 (\nabla\nabla)_*^*\phi \oplus (-1)(\pa\phi)^2 \Big]^N \right) 
   - \tau_{\mu\nu} \delta_B = 0
\,,
\label{tensorEoM}
\\[2pt]
 V(\phi) - V' (\phi)
  -  \sum_{N=1}^{N_{max}} \frac{\alpha_N}{2}\, 
\cT \left( \Big[ {\textstyle\frac12} \cR_{**}^{**}
    \oplus 2 (\nabla\nabla)_*^*\phi \oplus (-1)(\pa\phi)^2 \Big]^N \right) 
  - \tau_\phi \delta_B = 0
\,,
\label{scalarEoM}
\end{eqnarray} 
where the brane localized terms, $\tau_{\mu\nu}$ and $\tau_\phi$, 
are calculated from the brane Lagrangian $\cL_B$, namely
\begin{equation}
\tau_{\mu\nu}
= 
h_{\mu\nu} \, \cL_B - 2 \frac{\delta \cL_B}{\delta h^{\mu\nu}}
\,, 
\qquad\qquad
\tau_\phi
= 
\cL_B - \frac{\delta \cL_B}{\delta \phi} 
\,. 
\label{taus}
\end{equation}
In the tensor equation of motion (\ref{tensorEoM}) we introduced 
another generalization of the standard trace given by 
\begin{equation}
{\ocT}_\mu^{\,\nu} (M) = 
\delta_{\mu \rho_1 \rho_2 \cdots \rho_m}^{\nu \sigma_1 \sigma_2 \cdots \sigma_m}
    M^{\rho_1 \rho_2 \cdots \rho_m}{}_{\sigma_1 \sigma_2 \cdots \sigma_m}
\,.
\label{cTbar}
\end{equation}
It maps arbitrary rank $(m,m)$ tensors into rank $(1,1)$ 
ones\footnote{
For $m=d$ the generalized Kronecker delta  
$\delta_{\mu\rho_1 \rho_2 \cdots \rho_m}^{\nu\sigma_1 \sigma_2 \cdots \sigma_m}$
should be replaced in (\ref{cTbar}) 
with the following combination:
$
\delta_\mu^\nu
\delta_{\rho_1 \rho_2 \cdots \rho_m}^{\sigma_1 \sigma_2 \cdots \sigma_m}
-
\delta_{\,\mu}^{\sigma_1}
\delta_{\rho_1 \rho_2 \cdots \rho_m}^{\,\nu\;\sigma_2 \cdots \sigma_m}
- \ldots -
\delta_{\,\mu}^{\sigma_m}
\delta_{\rho_1 \rho_2 \cdots \rho_m}^{\sigma_1 \sigma_2 \cdots\,\nu}
$.
}.

The main goal of the present work is to find the effective 
$(d-1)$-dimensional equations of motion at the brane located 
at the support of $\delta_B$. 
The first step in this direction is to identify parts 
``parallel'' and ``perpendicular'' to the brane in all relevant 
tensors.
In order to achieve this in a covariant way, we start with introducing 
a vector field $n^\mu$ normalized to 1 and perpendicular to the 
brane at its position.
The choice of $n^\mu$ is not unique, due to the freedom in the 
bulk and the sign ambiguity at the brane. 
However, as we will see later, the effective brane equations 
of motion are unique.  
With any such a vector field $n^\mu$ we define the metric 
$h_{\mu\nu}=g_{\mu\nu}-n_\mu n_\nu$. 
This expression holds throughout the 
$d$-dimensional space-time but, restricted to the brane position, 
it just yields the metric induced on the brane. 
Subsequently, we divide the $d$-dimensional tensors into parts parallel 
and perpendicular to the vector field $n^\mu$ as
\begin{eqnarray} 
\cR^{\nu\sigma}_{\mu\rho} 
\,=\,
&&\hspace{-18pt}
R^{\nu\sigma}_{\mu\rho} - 2 \, K^\nu_{[\mu} K^\sigma_{\rho]} 
- 4\, n^{[\nu} D_{[\mu} K^{\sigma]}_{\rho]} 
- 4\, n_{[\mu} D^{[\nu} K^{\sigma]}_{\rho]} 
  - 4\, n^{[\nu} n_{[\mu} \Ln K^{\sigma]}_{\rho]} 
+ 4\, n^{[\nu} n_{[\mu} (KK)^{\sigma]}_{\rho]} 
\,, \quad
\label{projectedR}
\\[2pt]
\nabla^\nu\nabla_\mu\phi 
\,=\,
&&\hspace{-18pt}
D^\nu D_\mu\phi + K^\nu_\mu\Ln\phi 
+ n^\nu D_\mu \Ln\phi + n_\mu \, D^\nu \Ln\phi  
- n^\nu K_\mu^\sigma D_\sigma\phi - n_\mu \, K^\nu_\sigma D^\sigma\phi 
\nn[2pt]
&&\hspace{-18pt}
+ \, n^\nu n_\mu \left( \Ln^2 \phi - a^{\lambda} \nabla_{\lambda} \phi \right) 
,
\label{projectedBox}
\\[2pt]
(\pa\phi)^2 
\,=\,
&&\hspace{-18pt}
(D\phi)^2 + (\Ln\phi)^2
\,,
\label{projectedDD}
\end{eqnarray}
where $K_{\mu\nu}=\frac12\Ln h_{\mu\nu}$ is the extrinsic curvature of 
hypersurfaces orthogonal to $n^\mu$, $\Ln$ is the Lie 
derivative\footnote{
For a given tensor $M^{\rho_1\rho_2\cdots\rho_m}_{\sigma_1\sigma_2\cdots\sigma_l}$ 
and arbitrary direction $v^\mu$, 
the Lie derivative along $v^\mu$ is defined as follows:
$\pounds_v  \, M^{\rho_1\rho_2\cdots\rho_m}_{\sigma_1\sigma_2\cdots\sigma_l}
 = v^\lambda \nabla_\lambda M^{\rho_1\rho_2\cdots\rho_m}_{\sigma_1\sigma_2\cdots\sigma_l}
- \sum_{i=1}^m M^{\rho_1\rho_2\cdots\lambda\cdots\rho_m}
    _{\sigma_1\sigma_2\cdot\cdot\cdots\cdots\sigma_l}
  \nabla_\lambda v^{\rho_i}
+ \sum_{j=1}^l M^{\rho_1\rho_2\cdots\cdot\cdot\cdots\rho_m}
    _{\sigma_1\sigma_2\cdots\lambda\cdots\sigma_l}
  \nabla_{\sigma_j} v^\lambda$, 
thus e.g.\ $\Ln\phi=n^\lambda \nabla_\lambda \phi$.
} 
along $n^\mu$, and $a^{\lambda}\equiv n^\rho \nabla_\rho n^\lambda$. 
It is important to distinguish between the $d$-dimensional 
tensors associated with the full metric $g_{\mu\nu}$, 
namely the Riemann tensor $\cR^{\mu\nu}_{\rho\sigma}$ 
and the covariant derivative $\nabla_\mu$, and the $(d-1)$-dimensional 
tensors associated with the metric $h_{\mu\nu}$ restricted to the 
brane position, i.e.\ $R^{\mu\nu}_{\rho\sigma}$  and $D_\mu$.

The right hand sides of eqs.\ (\ref{projectedR})-(\ref{projectedDD}) 
are expressed almost entirely in terms of the vector field $n^\mu$, 
Lie derivatives along $n^\mu$ and brane quantities orthogonal to  
$n^\mu$, i.e.\ $R^{\mu\nu}_{\rho\sigma}$, $K_{\mu\nu}$ and $D_\mu$. 
The term 
$a^\lambda\nabla_\lambda\phi=n^\rho(\nabla_\rho n^\lambda)(\nabla_\lambda\phi)$,
containing the $d$-dimensional covariant derivatives,
is the only exception. However, 
as we will see later, the derivation procedure for 
the effective gravitational equations at the brane will be 
constructed in such a way, that $a^\lambda\nabla_\lambda\phi$  
will not appear in our final results.

The Riemann tensor $\cR_{\mu\nu}^{\rho\sigma}$
and the tensor of the second covariant derivative of the dilaton 
$\nabla^\nu\nabla_\mu\phi$ appear in the  bulk equations of motion 
(\ref{tensorEoM}) and (\ref{scalarEoM})
exclusively under the generalized traces (\ref{cT}) and (\ref{cTbar}), 
which involve full 
anti-symmetrization in all covariant and contravariant indices. 
Hence, the projection equations (\ref{projectedR}) and (\ref{projectedBox}) 
can be simplified when used under those traces, namely
\begin{eqnarray} \hspace{-0.5cm}
\cR_{**}^{**} &\to& R_{**}^{**} - 2 K^*_* K^*_*
  - 4 (nn)_*^* \big( \Ln K_*^* - (KK)^*_* \big) - 8 (nD)_*^* K_*^*
\,,
\label{projectedRast}
\\[2pt] \hspace{-0.5cm}
(\nabla\nabla)_*^* \phi &\to& \big[ (DD)_*^* \phi + K_*^* \Ln \phi \big]
  + (nn)_*^* \left( \Ln^2 \phi - a^\lambda \nabla_\lambda \phi \right)
  + 2 \big[ (nD)_*^* \Ln \phi - (nKD)_*^* \phi \big] 
\,,
\label{projectedBoxast}
\end{eqnarray}
where in order to make the formulae more compact we introduced  
the following notation:
$ (nn)_*^* \equiv n_* n^* $, $ (\nabla\nabla)_*^* \equiv \nabla_* \nabla^* $, 
$ (DD)_*^* \equiv D_* D^* $, 
$ (KK)^*_* \equiv K^*_\lambda K^\lambda_* $, 
$ (nD)_*^* \equiv \frac12 \left( n_* D^* + n^* D_* \right) $ and
$ (nKD)_*^* \equiv \frac12 
\left( n_* K^*_\lambda D^\lambda + n^* K_*^\lambda D_\lambda \right) $.

Employing the dimensional reduction formulae (\ref{projectedRast}), 
(\ref{projectedBoxast}) and (\ref{projectedDD}), the bulk equations 
of motion (\ref{tensorEoM}) and (\ref{scalarEoM}) can be rewritten as
\begin{eqnarray}
 g_{\mu\nu} V(\phi)
  -  \sum_{N=1}^{N_{max}} \frac{\alpha_N}{2}\, {\ocT}_{\mu\nu} 
&&\!\!\!\!\!\!\!\!\!\!\bigg( \Big[
    \Big( {\textstyle\frac12} R_{**}^{**} \oplus 2 (DD)_*^* \phi 
\oplus(-1) (D\phi)^2 \Big)
    \oplus(-1) \big( K^*_* \oplus(-1) \Ln \phi \big) ^2 
\nn[2pt]
&&\!\!\!
    \oplus \, 2 \, (nn)_*^* \big( (KK)^*_* - \Ln K_*^* \big)
    \oplus 2 \, (nn)_*^* \left( \Ln^2 \phi 
- a^\lambda \nabla_\lambda \phi \right) 
\nn[2pt]
&&\!\!\!
    \oplus \, (-4) (nD)_*^* K_*^* \oplus 4 \big[ (nD)_*^* \Ln \phi 
- (nKD)_*^* \phi \big] \Big]^N \bigg) 
   -  \tau_{\mu\nu} \, \delta_B 
=0
\,,\qquad
\label{PtensorEoM}
%\end{eqnarray}
%
%
%\begin{eqnarray}
\\[2pt]
 V(\phi) - V' (\phi)
  -  \sum_{N=1}^{N_{max}} \frac{\alpha_N}{2}\, \cT 
&&\!\!\!\!\!\!\!\!\!\!\bigg( \Big[
    \Big( {\textstyle\frac12} R_{**}^{**} \oplus 2 (DD)_*^* \phi 
\oplus(-1) (D\phi)^2 \Big)
    \oplus(-1) \big( K^*_* \oplus(-1) \Ln \phi \big) ^2 
\nn[2pt]
&&\!\!\!
    \oplus \, 2 \, (nn)_*^* \big( (KK)^*_*  - \Ln K_*^* \big)
    \oplus 2 \, (nn)_*^* \left( \Ln^2 \phi 
- a^\lambda \nabla_\lambda \phi \right) 
\nn[2pt]
&&\!\!\!
    \oplus \, (-4) (nD)_*^* K_*^* \oplus 4 \big[ (nD)_*^* \Ln \phi 
- (nKD)_*^* \phi \big] \Big]^N \bigg) 
   -  \tau_\phi \, \delta_B 
=0
\,.\qquad
\label{PscalarEoM}
\end{eqnarray}
These equations remain defined and valid throughout the full 
$d$-dimensional space-time. They are smooth in the bulk, 
but contain discontinuous and 
singular (distribution-like) terms localized at the brane. 
The effective brane equations of motion can be obtained if  
the information from both smooth 
and singular parts of (\ref{PtensorEoM}) and 
(\ref{PscalarEoM}) is properly taken into account at the position of the brane.

%%%%%%%%%%%%%%%%%%%%%%%%%%%%%%%%%%%%%%%%%%%%%%%%%%%%%%%%%%%%%%%%
\section{ Junction conditions }
\label{section_JC}
%%%%%%%%%%%%%%%%%%%%%%%%%%%%%%%%%%%%%%%%%%%%%%%%%%%%%%%%%%%%%%%%

In order to find the junction conditions which have to be fulfilled at 
the brane, we shall 
define a 1-dimensional integration in the direction perpendicular 
to the brane. In order to achieve this, we will employ a special 
family of integral 
curves associated with the vector field $n^\mu$. Specifically, 
let us define the curves 
 $\gamma^\mu(\lambda)$ by requiring the vector field 
$n^\mu$  to be at each point tangent to $\gamma^\mu$:
$({\rm d}\gamma^\mu/{\rm d}\lambda)(\lambda)=n^\mu(\gamma^\nu(\lambda))$. 
We choose the parameterization of these curves 
in such a way that each of them crosses the brane at 
the point $\gamma^\mu(0)$.  
For a scalar field $f(x^\mu)$, at each point $x^\mu$, we consider the 
following integral:
\begin{equation}
\int_{\Gamma^\mu_x(\lambda_1,\lambda_2)}
f(x^\mu)
=
\int_{\lambda_1}^{\lambda_2}f\big(\gamma^\mu_x(\lambda)\big){\rm d}\lambda
\,,
\label{int_lambda}
\end{equation}
where 
$\Gamma^\mu_x(\lambda_1,\lambda_2)
=
\left\{\gamma^\mu_x(\lambda)|\lambda\in[\lambda_1,\lambda_2]\right\}$ 
is a part of the curve $\gamma^\mu_x(\lambda)$ crossing the brane 
at the point $x^\mu$.
This integration is ``inverse'' to the Lie 
derivative\footnote{
From now on whenever we refer to a Lie derivative, it should be 
understood as the Lie derivative along the vector field $n^\mu$.
} 
in the following sense:
\begin{equation}
\int_{\Gamma^\mu_x(\lambda_1,\lambda_2)} \Ln f(x^\mu)
= f\big(\gamma^\mu_x(\lambda_2)\big)-f\big(\gamma^\mu_x(\lambda_1)\big)
\,.
\label{intLie}
\end{equation}
This integral can be generalized to the case of arbitrary tensor 
integrands by applying appropriate pullback and pushforward 
operations along the family of curves $\gamma^\mu(\lambda)$.

Junction conditions for a given point $x_0^\mu$ on the brane
are obtained by integrating the $d$-dimensional equations of motion 
along (an infinitesimal part of) the curve $\gamma^\mu_{x_0}(\lambda)$
going through $x^\mu_0$:
\begin{equation}
\int_{\perp B} F\left(x^\mu_0\right)
\equiv
\lim_{\epsilon\to0}
\int_{\Gamma^\mu_{x_0}(-\epsilon,+\epsilon)}
F\left({x^\mu_0}\right) 
,
\label{int_perp_B}
\end{equation}
where $F(x^\mu)$ is the scalar or the tensor given by the left hand sides of 
the bulk equations of motion (\ref{PtensorEoM}) and (\ref{PscalarEoM}), 
respectively\footnote{
Integral (\ref{int_perp_B}) is well defined for any 
point of the space-time. However, away from the brane the 
integrands are smooth and the integral vanishes in the $\epsilon\to0$ limit.  
Non-trivial results are obtained only on the 
brane, where there are localized sources $\tau_{\mu\nu}$ and 
$\tau_\phi$. }.

Only certain terms in the $d$-dimensional equations of motions
will not yield zero 
during the above infinitesimal across-the-brane integration, 
namely the brane contributions involving $\tau_{\mu\nu}$ and $\tau_\phi$, 
explicitly proportional to $\delta_B$, and the terms containing 
Lie derivatives (i.e.\ $\Ln^2\phi$ and $\Ln K_{\mu\nu}$)
of quantities which can be discontinuous at the brane 
($\Ln\phi$ and $K_{\mu\nu}$, respectively).
Hence, the relevant parts of the bulk equations of motions 
 (\ref{PtensorEoM}) and (\ref{PscalarEoM}) yield  
\begin{eqnarray} 
\sum_{N=1}^{N_{max}} \alpha_N N \int_{\perp B} {\ocT}_{\mu\nu} 
\bigg( (nn)_*^* \,
  \Ln \big( K_*^* \oplus (-1) \Ln \phi \big)
\hspace{220pt}&&
\nn[2pt]
  \cdot \Big[ 
\Big( {\textstyle\frac12} R_{**}^{**} \oplus 2 (DD)_*^* \phi 
\oplus(-1) (D\phi)^2 \Big)
\oplus(-1) \big( K^*_* \oplus(-1) \Ln \phi \big) ^2  \Big]^{N-1} \bigg)
=
&&\hspace{-15pt}
\tau_{\mu\nu}
\,,\qquad
\label{tensorJCint}
\\[2pt]
\sum_{N=1}^{N_{max}} \alpha_N N \int_{\perp B} \cT \bigg(
  \Ln \big( K_*^* \oplus (-1) \Ln \phi \big) 
\hspace{260pt}&&
\nn[2pt]
  \cdot \Big[ 
\Big( {\textstyle\frac12} R_{**}^{**} \oplus 2 (DD)_*^* \phi 
\oplus(-1) (D\phi)^2 \Big)
\oplus(-1) \big( K^*_* \oplus(-1) \Ln \phi \big) ^2  \Big]^{N-1} \bigg)
=
&&\hspace{-15pt}
\tau_\phi  
\,.
\label{scalarJCint}
\end{eqnarray} 
The integration of the terms containing the brane localized sources
$\tau_{\mu\nu}$ and $\tau_\phi$
was easy due to the obvious (defining) properties of the Dirac 
delta distribution $\delta_B$. The remaining integrations require  
a more careful treatment. The integrands involve products 
of the distribution-like objects, $\Ln K_{\mu\nu}$ or $\Ln^2\phi$, and 
the potentially discontinuous objects, $K_{\mu\nu}$ or $\Ln\phi$. Strictly 
speaking, such integrals have no mathematically unambiguous meaning, 
as distributions are defined by their integrals with smooth functions. 
Therefore, we need some kind of regularization to deal with 
such terms. Fortunately, 
there is an obvious way to regularize and calculate the integrals 
in eqs.\ (\ref{tensorJCint}) and (\ref{scalarJCint}). 
Let us first consider terms of the form $sf^k\Ln f$, where $s$ is 
smooth, whereas $f$ may be discontinuous at the brane. 
Employing the formula (\ref{intLie}) and the Leibniz rule for the Lie 
derivative we get
\begin{equation}
\int_{\perp B} sf^k \Ln f
=
{\textstyle\frac{1}{k+1}}
\int_{\perp B} s \Ln \left(f^{k+1}\right)
=
{\textstyle\frac{1}{k+1}}
\int_{\perp B} \Ln \left(sf^{k+1}\right)
=
{\textstyle\frac{1}{k+1}}
[sf^{k+1}]_\pm
\,.
\label{regularization}
\end{equation}
The second equality follows from the fact that the integral of 
$(\Ln s)f^{k+1}$ vanishes in the $\epsilon\to0$ limit. 
The square bracket with the $\pm$ subscript was introduced to 
denote a jump in the value of some quantity when crossing the brane:
\begin{equation}
[f(x^\mu_0)]_\pm
=
[f(x^\mu_0)]_+-[f(x^\mu_0)]_-
\,,
\label{bracket_pm}
\end{equation}
while the square brackets with the subscripts $+$ and $-$ denote
the limits of a given bulk quantity when approaching the brane
from the ``$+$'' and the ``$-$'' sides, respectively:
\begin{equation}
[f(x^\mu_0)]_{+(-)}
=
\lim_{\epsilon\to0^{+(-)}}f
\left(\gamma^\mu_{x_0}(\epsilon)\right)
.
\label{bracket_+(-)}
\end{equation}

We intend to apply the regularization (\ref{regularization}) to the 
integrals in eqs.\ (\ref{tensorJCint}) and (\ref{scalarJCint}). 
It is instructive to start from considering a simple example 
of such a calculation:
\begin{eqnarray}
&&
\int_{\perp B}\cT\Big(\big[K^*_*\oplus(-1)\Ln\phi\big]
\Ln\big[K^*_*\oplus(-1)\Ln\phi\big]\Big) 
\nn[2pt]
&&\qquad\quad=
\int_{\perp B}\Big[
\delta_{\mu\rho}^{\nu\sigma}K^\mu_\nu\Ln K^\rho_\sigma
-\delta_\mu^\nu \left(\Ln K^\mu_\nu\Ln\phi + K^\mu_\nu \Ln^2\phi\right)
+\Ln^2\phi\Ln\phi 
\Big]
\nn[2pt]
&&\qquad\quad=
{\textstyle\frac12}\,\delta_{\mu\rho}^{\nu\sigma}
\int_{\perp B}\Ln\left(K^\mu_\nu K^\rho_\sigma\right)
-\delta_\mu^\nu \int_{\perp B}\Ln\left(K^\mu_\nu\Ln\phi\right)
+{\textstyle\frac12}\int_{\perp B}\Ln\left(\Ln\phi\right)^2 
\nn[2pt]
&&\qquad\quad=
{\textstyle\frac12}\,\delta_{\mu\rho}^{\nu\sigma}
\left[K^\mu_\nu K^\rho_\sigma\right]_\pm
-\delta_\mu^\nu \left[K^\mu_\nu\Ln\phi\right]_\pm
+{\textstyle\frac12}\left[(\Ln\phi)^2\right]_\pm
=\Big[\cT\Big({\textstyle\frac12}\left\{K^*_*\oplus(-1)\Ln\phi\right\}^2\Big)
\Big]_\pm
\,.\qquad
\nonumber
\end{eqnarray}
The same result is obtained if the regularization (\ref{regularization}) is used when
treating the formal sums of tensors of different ranks,  
under the generalized traces $\cT$ and ${\ocT}_{\mu\nu}$, 
as ordinary functions. Performing similar calculations for all 
terms present in (\ref{tensorJCint}) and (\ref{scalarJCint}), 
we obtain the junction conditions as
\begin{eqnarray}
\sum_{N=1}^{N_{max}} \alpha_N N \sum_{k=0}^{N-1} 
\frac{(-1)^k(N-1)!}{(2k+1)k!(N-1-k)!} 
\hspace{260pt}&&
\nn[2pt]
   \cdot \left[ {\ocT}_{\mu\nu} \left( (nn)^*_*
    \Big[ {\textstyle\frac12} R_{**}^{**} \oplus 2 (DD)_*^* \phi 
\oplus(-1) (D\phi)^2 \Big]^{N-1-k}
    \Big( K_*^* \oplus(-1) \Ln \phi \Big)^{2k+1} \right) \right]_\pm 
=
&&\hspace{-15pt}
\tau_{\mu\nu}
\,,\qquad
\label{tensorJC}
\\[2pt]
\sum_{N=1}^{N_{max}} \alpha_N N \sum_{k=0}^{N-1}
  \frac{(-1)^k(N-1)!}{(2k+1)k!(N-1-k)!} 
\hspace{260pt}&&
\nn[2pt]
  \cdot \left[ \cT \left( \Big[ {\textstyle\frac12} R_{**}^{**} 
\oplus 2 (DD)_*^* \phi \oplus(-1) (D\phi)^2 \Big]^{N-1-k}
  \Big( K_*^* \oplus(-1) \Ln \phi \Big)^{2k+1} \right) \right]_\pm 
= 
&&\hspace{-15pt}
\tau_\phi
\,.\quad
\label{scalarJC}
\end{eqnarray}
These formulae can be rewritten in a slightly 
(but qualitatively) different form 
in the case of theories with the bulk $\Z2$ symmetry  
and the brane located at the $\Z2$ symmetry fixed point. 
Then, for any $\Z2$-odd quantity $f$,  
$[f]_-=-[f]_+$ and $[f]_\pm$ can be replaced with $2\,[f]_+$.

A comment on distinguishing both sides of the brane 
is in order here. We started our construction with the 
vector field $n^\mu$, which at the brane is normal to it. 
The vector field $-n^\mu$ has the same property. 
Replacing $n^\mu$ with $-n^\mu$ (which 
corresponds to interchanging the ``$+$'' and the ``$-$'' 
sides of the brane) we have to change the sign of $\lambda$ 
in the family of curves $\gamma^\mu_{x}(\lambda)$ used 
in (\ref{int_perp_B}) to define the integral $\int_{\perp B}$\,. 
From (\ref{bracket_pm}) it follows that this in turn changes 
the sign of $[\cdots]_\pm$\,. The vector field $n^\mu$ enters 
linearly also the definitions of the Lie derivative $\Ln$ 
and the extrinsic curvature $K_{\mu\nu}$. Hence, the 
combination $\left(K_*^* \oplus(-1) \Ln \phi \right)$
changes sign together with $n^\mu$. Observe that 
only odd powers of this combination appear in the junction conditions 
(\ref{tensorJC}) and (\ref{scalarJC}). Thus, the change of 
sign in the definition (\ref{bracket_pm}) of $[\cdots]_\pm$ is compensated 
by the change of sign of  $\Ln\phi$ and $K_{\mu\nu}$.
The left hand sides of (\ref{tensorJC}) and (\ref{scalarJC}) 
do not depend on the sign of $n^\mu$ - as it should be, 
as the brane localized interactions (\ref{taus}) appearing on the 
right hand sides of these equations do not depend on which side
of the brane is called the ``$+$'' one.

The junction conditions (\ref{tensorJC}) and (\ref{scalarJC}) 
determine the jumps in the values of the extrinsic 
curvature, $K_{\mu\nu}$, and the Lie derivative of the scalar 
field, $\Ln\phi$, caused by the brane localized interactions 
described by $\tau_{\mu\nu}$ and $\tau_\phi$. 
Unfortunately, the solutions for $[K_{\mu\nu}]_\pm$ and $[\Ln\phi]_\pm$ 
(or $[K_{\mu\nu}]_+$ and $[\Ln\phi]_+$ in the case of 
$\Z2$-symmetric models) can be found explicitly only 
in some simple cases, for example when $N_{\rm max}=1$ 
or when we are considering solutions which are highly symmetric.

%%%%%%%%%%%%%%%%%%%%%%%%%%%%%%%%%%%%%%%%%%%%%%%%%%%%%%%%%%%%%%%%
\section{ Effective brane equations of motion }
\label{section_effectiveEoM}
%%%%%%%%%%%%%%%%%%%%%%%%%%%%%%%%%%%%%%%%%%%%%%%%%%%%%%%%%%%%%%%%

%%%%%%%%%%%%%%%%%%%%%%%%%%%%%%%%
\subsection{ Definitions }
\label{subsection_definitions}
%%%%%%%%%%%%%%%%%%%%%%%%%%%%%%%%

In the previous section we found the junction conditions 
(\ref{tensorJC}) and (\ref{scalarJC}) which have to be fulfilled 
at the brane. Presently we would like to obtain $(d-1)$-dimensional 
equations which we could call the ``effective brane equations of 
motion''. First of all, a precise definition of such effective 
equations of motion at the brane has to be given. There are two 
obvious properties of these equations: First, they must follow from 
the full $d$-dimensional equations of motion (\ref{PtensorEoM}) 
and (\ref{PscalarEoM}). Second, they should describe the 
behavior of the quantities defined exactly on the brane, or 
infinitesimally close to it. We can obtain such equations in a very 
similar way to that employed in  the junction conditions derivation. 
Specifically, for each point $x_0^\mu$ on the brane, we integrate 
the full $d$-dimensional equations of motion (\ref{PtensorEoM}) 
and (\ref{PscalarEoM}) over some infinitesimal interval ``perpendicular'' 
to the brane. However, the integration (\ref{int_lambda}) has to 
be generalized in order to obtain something different than 
just the junction conditions (\ref{tensorJC}) and (\ref{scalarJC}). 
An obvious way to generalize the integral (\ref{int_lambda}) is to use 
a weight function $w(\lambda)$:
\begin{equation}
\int_{\lambda_1}^{\lambda_2}f\big(\gamma^\mu_x(\lambda)\big)
\,w(\lambda)\,{\rm d}\lambda
\,,
\label{int_w}
\end{equation}
where $\gamma^\mu_{x_0}(\lambda)$ is again the integral curve of the 
vector field $n^\mu$ intersecting the brane at the point $x^\mu_0$. 
In fact, we need a family of weight functions $w_\epsilon(\lambda)$ 
such that the support of $w_\epsilon(\lambda)$ is 
included in the interval $(-\epsilon,+\epsilon)$. Any such a smooth 
function can be written as the following sum:
\begin{equation}
w_\epsilon(\lambda)
=
w_\epsilon^{(0)}+w_\epsilon^{(-)}(\lambda)+w_\epsilon^{(+)}(\lambda)
\,,
\end{equation}
where $w_\epsilon^{(0)}$ is a constant (finite in the $\epsilon\to0$ 
limit), $w_\epsilon^{(-)}(\lambda)$ has the support in $(-\epsilon,0)$ 
and $w_\epsilon^{(+)}(\lambda)$ has the support in $(0,+\epsilon)$.
The brane equations of motion are then obtained by integrating the 
$d$-dimensional equations of motion (\ref{PtensorEoM}), (\ref{PscalarEoM})
like in (\ref{int_w}) and taking the limit of 
$\epsilon\to0$, namely
\begin{equation}
\lim_{\epsilon\to0}
\int_{\Gamma^\mu_{x_0}(-\epsilon,+\epsilon)}
F\big(\gamma^\mu_{x_0}(\lambda)\big) w_\epsilon(\lambda)\,{\rm d}\lambda
=
c_0\int_{\perp B}F(x^\mu_0)
+c_-\lim_{\epsilon\to0^-}F(x^\mu_0+\epsilon\,n^\mu)
+c_+
\lim_{\epsilon\to0^+}F(x^\mu_0+\epsilon\,n^\mu)
\,,
\label{3_brane_eqs}
\end{equation}
where the coefficients $c$ depend on the chosen $w_\epsilon$:
\begin{equation}
c_0=\lim_{\epsilon\to0}w_\epsilon^{(0)}
\,,\qquad\qquad
c_-=\lim_{\epsilon\to0}\int_{-\epsilon}^{0}w_\epsilon^{(-)}(\lambda){\rm d}\lambda
\,,\qquad\qquad
c_+=\lim_{\epsilon\to0}\int_{0}^{+\epsilon}w_\epsilon^{(+)}(\lambda){\rm d}\lambda
\,.
\end{equation}
By adopting different $w_\epsilon$ we obtain three independent 
$(d-1)$-dimensional equations. The first term on the r.h.s.\ 
of (\ref{3_brane_eqs}) is proportional to the corresponding junction 
condition discussed already in the previous section. The remaining 
two terms are proportional to two directional limits of the bulk equations 
(i.e.\ the limits of the bulk equations when approaching the brane
from the ``$+$'' or the ``$-$'' side).

None of these equations can be called an effective brane equation of 
motion. The directional limits of the bulk equations have no explicit 
dependence on the brane localized quantities (\ref{taus}). 
The junction conditions do depend on the brane localized quantities,  
like the brane energy momentum tensor $\tau_{\mu\nu}$, 
but they determine the jumps in the
bulk quantities values and not the dynamics of the brane quantities. 
However, one should not draw the conclusion that it is not possible 
to define any effective brane equations of motion, as
there are more $d$-dimensional bulk equations than are needed for 
a $(d-1)$-dimensional brane gravity. In addition, one should remember 
that the metric tensor and the scalar field are continuous at the brane. 
The effective equations of motion at the brane are obtained by 
combining all available information: the junction conditions, 
the directional limits of all bulk equations and the continuity conditions.  
This will be explained in more detail in the next subsection - employing 
our model with the bulk equations of motion 
(\ref{PtensorEoM}) and (\ref{PscalarEoM}).

%%%%%%%%%%%%%%%%%%%%%%%%%%%%%%%%
\subsection{ Construction -- general case}
\label{subsection_general_case}
%%%%%%%%%%%%%%%%%%%%%%%%%%%%%%%%

We start with projecting the bulk tensor equation of motion 
(\ref{PtensorEoM}) on the brane hypersurface and/or 
on the normal vector field $n^{\mu}$. Multiplying 
(\ref{PtensorEoM}) by $h_\rho^\mu h_\sigma^\nu$, using 
the fact that $n^\mu$ is orthogonal to all brane directions,
and taking appropriate limits of the bulk fields, we get the 
following two equations (one for each side of the brane):
\begin{eqnarray}
\bigg[
h_{\rho\sigma} V(\phi) 
-\!\sum_{N=1}^{N_{max}} 
\frac{\alpha_N}{2}\, h_\rho^\mu h_\sigma^\nu \, {\ocT}_{\mu\nu}
\bigg( \cM^N
    \oplus\, 2N 
\Big[\big( (KK)^*_* - \Ln K_*^* \big)
 \oplus \left( \Ln^2 \phi - a^\lambda \nabla_\lambda \phi \right) \Big]
\cM^{N-1}
&&
\nn[2pt]
\oplus\,8N(1-N)\,
\cN
\cM^{N-2}
\bigg)
\bigg]_{-(+)}
\!=0\,
,
\hspace{94pt}&&
\label{ThhEoM}
\end{eqnarray}
where we introduced symbols $\cM$ and $\cN$ to denote the following 
formal combinations of the tensors of different ranks:
\begin{eqnarray}
\cM
=
&&\hspace{-18pt}
{\textstyle\frac12} R_{**}^{**} \oplus 2 (DD)_*^* \phi \oplus(-1) (D\phi)^2 
    \oplus(-1) \big( K^*_* \oplus(-1) \Ln \phi \big) ^2 
\,,
\label{cM}
\\[2pt]
\cN
=
&&\hspace{-18pt}
\left[ D_* K_*^* 
\oplus  \left( K_*^\lambda D_\lambda\phi - D_* \Ln\phi \right) \right]
\left[ D^* K_*^* 
\oplus  \left( K^*_\lambda D^\lambda\phi - D^* \Ln\phi \right) \right] 
,
\label{cN}
\end{eqnarray}
which will appear under the generalized traces 
$\cT$ and ${\ocT}_{\mu\nu}$ in numerous formulae. 
The index structure of $\cN$ is different than that of all 
other objects appearing under $\cT$ and ${\ocT}_{\mu\nu}$.
So far all such objects had natural pairs (covariant-contravariant) 
of indices - corresponding 
to the pairs of indices present in the generalized 
Kronecker delta (\ref{delta}). In $\cN$ there is one unpaired 
covariant index in the first square bracket in (\ref{cN}) and one 
unpaired contravariant index in the second square bracket. 
It should be understood that these two indices form a pair under  
$\cT$ and ${\ocT}_{\mu\nu}$. Thus, we have for example
\begin{eqnarray}
\cT(\cN)
=
\delta_{\rho_1\rho_2\rho_3}^{\sigma_1\sigma_2\sigma_3}
\left(D_{\sigma_1}K_{\sigma_2}^{\rho_2}\right)
\left(D^{\rho_1}K_{\sigma_3}^{\rho_3}\right)
+
\delta_{\rho_1\rho_2}^{\sigma_1\sigma_2}
\left(D_{\sigma_1}K_{\sigma_2}^{\rho_2}\right)
\left(K^{\rho_1}_{\lambda}D^{\lambda}\phi - D^{\rho_1}\Ln\phi\right)
+\ldots 
\nonumber
\end{eqnarray}

The bulk dilaton equation of motion (\ref{PscalarEoM}) is very similar
to the projection (\ref{ThhEoM}) of the bulk tensor equation of motion 
on the brane (i.e. obtained by contracting with $h_\rho^\mu h_\sigma^\nu$) and reads
\begin{eqnarray}
\bigg[
V(\phi) - V'(\phi) 
-\!\sum_{N=1}^{N_{max}} 
\frac{\alpha_N}{2}\, \cT
\bigg( \cM^N
     \oplus\, 2N 
\Big[\big( (KK)^*_* - \Ln K_*^* \big)
 \oplus \left( \Ln^2 \phi - a^\lambda \nabla_\lambda \phi \right) \Big]
\cM^{N-1}
\nn[2pt]
\oplus\,8N(1-N)\,
\cN
\cM^{N-2}
\bigg)
\bigg]_{-(+)}
\!=0
\,.
\hspace{94pt}&&
\label{WEoM}
\end{eqnarray}
The contractions of the bulk tensor equation of motion
(\ref{PtensorEoM}) with $h_\rho^\mu n^\nu$ 
and  $n^\mu n^\nu$ are given by, respectively,
\begin{eqnarray}
\left[
\sum_{N=1}^{N_{max}} \alpha_N \, {\ocT}_{\rho\sigma}
\bigg( N \Big[
D^\sigma K_*^* 
\oplus \left( K^\sigma_\lambda D^\lambda \phi - D^\sigma \Ln \phi \right)
\Big]
\cM^{N-1}
\bigg)
\right]_{-(+)}=0
\,,&&
\label{ThnEoM}
\\[2pt]
\left[V(\phi)
-\sum_{N=1}^{N_{max}} \frac{\alpha_N}{2}\,
 \cT \Big( \cM^N \Big)
\right]_{-(+)}=0
\,.&&
\label{TnnEoM}
\end{eqnarray}

Before discussing the properties of the above equations of motion,
let us rewrite the tensor equation (\ref{ThhEoM}) in a somewhat 
different form. One of the reasons is that we would like to 
remove from our equations all terms containing the quantity 
$a^\lambda\nabla_\lambda\phi$, which is the only term being neither 
``parallel'', nor ``perpendicular'' to the brane. Another reason 
is that we would like to rewrite the tensor equation (\ref{ThhEoM}) in a form 
more suitable to compare our results with those presented so far 
in the literature. In order to achieve these goals, we conduct the 
following procedure: 
Using the decomposition (\ref{projectedR}) 
and the definition of 
the $d$-dimensional Weyl tensor $\cC_{\mu\nu\rho\sigma}$:
\begin{equation}
\cC_{\mu\nu\rho\sigma}
=
\cR_{\mu\nu\rho\sigma}
- {\textstyle\frac{2}{d-2}}
\Big(g_{\mu[\rho}\cR_{\sigma]\nu}-g_{\nu[\rho}\cR_{\sigma]\mu}\Big)
+{\textstyle\frac{2}{(d-1)(d-2)}}\,
g_{\mu[\rho} g_{\sigma]\nu}\cR
\,,
\end{equation}
we obtain the equation
\begin{eqnarray}
\Ln K_{\mu\nu} - (KK)_{\mu\nu}
 =
&&\hspace{-18pt} 
{\textstyle\frac{1}{d-1}} h_{\mu\nu} \big[ (h \Ln K) - (KK) \big]
  - {\textstyle\frac{1}{d-3}} \big[ R_{\mu\nu} - KK_{\mu\nu} 
+ (KK)_{\mu\nu} \big] 
\nn[2pt]
&&\hspace{-18pt}
+ {\textstyle\frac{1}{(d-1)(d-3)}} h_{\mu\nu} \big[ R - K^2 + (KK) \big]
  - {\textstyle\frac{d-2}{d-3}} E_{\mu\nu} 
\, ,
\label{LnK_E}
\end{eqnarray}
expressing the Lie derivative of the extrinsic curvature, $\Ln K_{\mu\nu}$,
as a function of its trace, $(h\Ln K)\equiv h^{\mu\nu}\Ln K_{\mu\nu}$, 
and the following projection of the $d$-dimensional Weyl tensor:
\begin{equation}
E_{\mu\nu}
=
n^\alpha h_\mu^\beta  n^\gamma h_\nu^\delta \;
\cC_{\alpha\beta\gamma\delta}
\,.
\label{E}
\end{equation}
After using eq.\ (\ref{LnK_E}), the tensor equation
(\ref{ThhEoM}) depends (explicitly) only on two second Lie
derivatives of the bulk fields, $(h\Ln K)$ and ${\Ln}^2\phi$, 
which appear always in the 
combinations \mbox{$[(h\Ln K)-(KK)]$} and 
\mbox{$[{\Ln}^2\phi- a^\lambda\nabla_\lambda\phi]$}. 
These two combinations can be calculated from a system of two linear 
algebraic equations consisting of the scalar equation (\ref{WEoM}) 
and the trace of the tensor equation (\ref{ThhEoM})
- both evaluated ``next to the brane''. 
The solution of this system of equations reads
\begin{eqnarray}
 (h \Ln K) - (KK)
=
&&\hspace{-18pt}
(d-1)
\frac{b_0 B_1 - b_1 B_0}{b_0 b_2 - b_1^2}
\,,
\label{TrLnK_sol}
\\[2pt]
\Ln^2\phi - a^\lambda\nabla_\lambda\phi
=
&&\hspace{-18pt}
\frac{b_1 B_1 - b_2 B_0}{b_0 b_2 - b_1^2}
\,,
\label{Ln2phi_sol}
\end{eqnarray}
where
\begin{eqnarray}
b_m
=
&&\hspace{-18pt}
\sum_{N=1}^{N_{\rm max}} \alpha_N N \,
\cT \Big( \left(h_*^*\right)^m \cM^{N-1} \Big) 
\,,
\label{bm}
\\[2pt]
B_m 
=
&&\hspace{-18pt}
\sum_{N=1}^{N_{\rm max}} \frac{\alpha_N}{2}\, 
\cT
\bigg( \! \left(h_*^*\right)^m  \Big[ \cM^N
 \oplus {\textstyle\frac{2N}{d-3}} \cP \cM^{N-1}
\oplus 8 N (1-N) \, \cN \cM^{N-2} \Big] 
\bigg)
\nn[2pt]
&&\hspace{-18pt}
-\,(d-1)^m \, V(\phi)+(1-m)\, V'(\phi)
\,,
\label{Bm}
\\[2pt]
\cP
=
&&\hspace{-18pt}
R_*^* - K K_*^* + (KK)_*^* 
- {\textstyle\frac{1}{d-1}} h_*^* \left[ R - K^2 + (KK) \right] + (d-2) E_*^* 
\,,
\label{cP}
\end{eqnarray} 
whereas $\cM$ and $\cN$ are defined in (\ref{cM}) and (\ref{cN}), 
respectively. Substituting eqs.\ (\ref{LnK_E}), (\ref{TrLnK_sol}) 
and (\ref{Ln2phi_sol}) into the projected tensor equation (\ref{ThhEoM}), 
we obtain a tensor equation of motion without explicit dependence 
on the second Lie derivatives of the bulk fields, namely
\begin{eqnarray}
\Bigg[
h_{\rho\sigma} V
-\!\sum_{N=1}^{N_{\rm max}} 
\frac{\alpha_N}{2}\, h^\mu_\rho h^\nu_\sigma \, {\ocT}_{\mu\nu}
\bigg( \cM^N
\oplus{\textstyle\frac{2N}{d-3}}\cP\cM^{N-1}
\oplus\,8N(1-N) \,\cN \cM^{N-2}
\hspace{72pt}
\nn[2pt]
    \oplus\, 2N 
\left[
h_*^* \frac{b_1 B_0 - b_0 B_1}{b_0 b_2 - b_1^2}
\oplus\frac{b_1 B_1 - b_2 B_0}{b_0 b_2 - b_1^2}
\right]
\cM^{N-1}
\bigg)\Bigg]_{-(+)}
=0
\,.
&&
\label{Thh_E}
\end{eqnarray}
It is useful to reformulate this equation further in order to 
rewrite it as a sum of the ordinary lowest order Einstein 
equation and some corrections. The first step is very simple, 
we just have to multiply (\ref{Thh_E}) by $(d-3)/(d-2)$ in order to get 
the usual normalization of the Ricci tensor. 
A second step is necessary, as the ratio of the coefficients 
in front of $h_{\rho\sigma}R$ and $R_{\rho\sigma}$ 
is different than the desired $-1/2$. To solve this problem 
we have to add a product of eq.\ (\ref{TnnEoM}) and 
the brane metric tensor $h_{\rho\sigma}$ with the appropriate 
coefficient. Finally, we obtain the following Einstein-like tensor equation
of motion:
\begin{eqnarray}
&& \hspace{-1.5cm}
\left[
{\textstyle\frac{(d-3)(2d-3)}{(d-1)(d-2)}}
h_{\rho\sigma} V
-{\textstyle\frac{d-3}{d-2}}\sum_{N=1}^{N_{\rm max}} 
\frac{\alpha_N}{2}\, h^\mu_\rho h^\nu_\sigma \, {\ocT}_{\mu\nu}
\bigg( \cM^N
\oplus{\textstyle\frac{2N}{d-3}}\cP\cM^{N-1}
\oplus\,8N(1-N) \,\cN \cM^{N-2}
\right.
%\hspace{-4pt}&&
\nn[2pt]
&& \hspace{-0.5cm}
\left. 
    \oplus\, 2N 
\left[
h_*^* \frac{b_1 B_0 - b_0 B_1}{b_0 b_2 - b_1^2}
\oplus\frac{b_1 B_1 - b_2 B_0}{b_0 b_2 - b_1^2}
\right]
\cM^{N-1}
\bigg)
-{\textstyle\frac{d-3}{d-1}}\sum_{N=1}^{N_{\rm max}} 
\frac{\alpha_N}{2}\,h_{\rho\sigma} \cT\Big(\cM^N\Big)
\right]_{-(+)}
=0
\,.
%\hspace{12pt}&&
\label{Thh_EE}
\end{eqnarray}
The above equation seems to be rather complicated, but, in fact, 
after calculating the generalized traces (\ref{cT}) and (\ref{cTbar}) it takes 
the most Einstein-like form we could obtain, as we shall
show explicitly for $N_{\rm max}=1,2$. 
Moreover, it does not depend on the Lie derivative of 
the extrinsic curvature, $\Ln K_{\mu\nu}$, 
and on the second Lie derivative of the dilaton, $\Ln^2\phi$. 
The whole dependence on these quantities, which cannot
be restricted by the junction conditions, is encoded 
in the projection (\ref{E}) of the bulk Weyl tensor on the 
brane, $E_{\mu\nu}$. This dependence is implicit via 
the quantities $B_m$ and $\cP$ defined in (\ref{Bm}) 
and (\ref{cP}), respectively.
Due to the explicit bulk contribution, represented by $E_{\mu\nu}$,
the Einstein-like equation of motion (\ref{Thh_EE})
does not form a closed system. 
Consequently, to describe fully the brane dynamics,
the bulk solution usually would have to be known.

All the equations we wrote down so far in this subsection 
are directional limits of the bulk equations of motion 
(or their combinations). In order to get any effective brane 
equations of motion we need some dependence on the brane 
sources (\ref{taus}). Such a dependence can be introduced by 
taking into account the junction conditions, which determine 
the jumps in the values of $K_{\mu\nu}$ and $\Ln\phi$,
i.e.\ $[K_{\mu\nu}]_\pm$ and $[\Ln\phi]_\pm$, in terms of the 
sources $\tau_{\mu\nu}$ and $\tau_\phi$. Let us now discuss 
what equations are obtained after substituting the 
solutions (not always known in a closed analytic form) 
of the junction conditions (\ref{tensorJC}) and (\ref{scalarJC}) 
into the bulk equations (\ref{ThhEoM}), (\ref{WEoM}), 
(\ref{ThnEoM}) and (\ref{TnnEoM})\footnote{
For this discussion it is better to use eqs.\ (\ref{ThhEoM}), 
(\ref{WEoM}) and (\ref{TnnEoM}), than the derived from them 
eqs.\ (\ref{Thh_E}) and (\ref{Thh_EE}).
}.

The first obvious observation is that eqs.\ 
(\ref{ThhEoM}) and (\ref{WEoM}) give no useful effective 
brane equations of motion, as they involve 
second derivatives of the brane metric and the 
dilaton field along the vector field $n^\mu$ normal 
to the brane, i.e.\ $\Ln K_{\mu\nu}$ and $\Ln^2\phi$, respectively. 
To show this more explicitly we consider 
the following construction:
We choose some brane sources $\tau_{\mu\nu}$ and $\tau_\phi$,
together with a metric $h_{\mu\nu}$ and a scalar field $\phi$ on 
the brane. More precisely, we choose $[h_{\mu\nu}]_+$ and 
$[\phi]_+$ (which is the same as $[h_{\mu\nu}]_-$ and $[\phi]_-$,
as the fields are continuous) at one of the two hypersurfaces 
infinitesimally close to the brane. Due to the junction conditions, 
the brane sources $\tau_{\mu\nu}$ and $\tau_\phi$ restrict (or in some 
cases determine) the first derivatives in the direction perpendicular 
to the brane, i.e.\ $[K_{\mu\nu}]_+$ and $[\Ln\phi]_+$ 
(or $[K_{\mu\nu}]_-$ and $[\Ln\phi]_-$) at the chosen hypersurface 
``next to the brane''. The values of these fields (i.e.\ 
$[h_{\mu\nu}]_+$ and $[\phi]_+$) and their first derivatives 
constitute simply the boundary conditions for the quasi-linear second 
order differential eqs.\ (\ref{ThhEoM}) and (\ref{WEoM}). 
The Cauchy-Kowalewski theorem tells us that such boundary
conditions problem can be solved at least in some neighborhood 
of the brane. The existence of the solutions does not require 
any relations between the brane fields and the brane 
sources.  Without any additional assumptions 
(e.g.\ about the symmetries of the bulk/brane solutions) 
the eqs.\ (\ref{ThhEoM}) and (\ref{WEoM}) and 
the junction conditions (\ref{tensorJC}) and (\ref{scalarJC}) 
do not yield any effective brane equations of motion, as
they are not sufficient to get any constraints on the brane fields 
$h_{\mu\nu}$ and $\phi$
in terms of the brane sources $\tau_{\mu\nu}$ and $\tau_\phi$.

From the above reasoning it is clear that in order to get 
any effective $(d-1)$-dimensional brane equations we need 
bulk equations of motion which do not involve the second 
derivatives perpendicular to the brane. Such equations,  
obtained from the $d$-dimensional tensor equation of motion 
by contracting at least one of its indices with the vector field 
$n^\mu$ normal to the brane, are given by (\ref{ThnEoM}) 
and (\ref{TnnEoM}). 
Let us start with the latter. The ``$+$'' side 
eq.\ (\ref{TnnEoM}) is just a consistency condition 
on the ``$+$'' side first Lie derivatives of the brane fields, 
$[K_{\mu\nu}]_+$ and $[\Ln\phi]_+$. Using the relations
\begin{equation}
[K_{\mu\nu}]_- = [K_{\mu\nu}]_+ - [K_{\mu\nu}]_\pm
\,,\qquad\qquad
[\Ln\phi]_- = [\Ln\phi]_+ - [\Ln\phi]_\pm 
\,,
\nonumber
\end{equation}
the ``$-$'' side eq.\ (\ref{TnnEoM}) becomes another 
consistency condition on $[K_{\mu\nu}]_+$ and $[\Ln\phi]_+$, 
this time depending on the brane sources $\tau_{\mu\nu}$ and $\tau_\phi$. 
It can be seen that 
in general, even in the simplest $N_{\rm max}=1$ case, 
these two consistency conditions obtained from 
eq.\ (\ref{TnnEoM}) can be fulfilled, thus eq.\ (\ref{TnnEoM}) 
does not lead to any effective brane equations of motion. 
The situation may change if some assumptions 
about the bulk are made, e.g.\  if the bulk $\Z2$ symmetry,
relating $[K_{\mu\nu}]_+$ with $[K_{\mu\nu}]_-$,
and thus with $[K_{\mu\nu}]_\pm$ and the brane sources 
$\tau_{\mu\nu}$ and $\tau_\phi$, is assumed. 
Such models will be discussed in the next subsection.

We have shown that in general eqs.\ (\ref{ThhEoM}), 
(\ref{WEoM}) and  (\ref{TnnEoM}) cannot be used to obtain 
any effective brane equations. 
What remains to discuss now, are two eqs.\
(\ref{ThnEoM}) - one for each side of the brane.  
They were obtained from the full $d$-dimensional 
tensor equation of motion (\ref{PtensorEoM}) by contracting one of the 
indices with $n^\mu$ and the other one with $h_{\mu\nu}$. The ``$+$'' 
side eq.\ (\ref{ThnEoM}) gives $(d-1)$ additional
consistency conditions on $[K_{\mu\nu}]_+$ and $[\Ln\phi]_+$. 
Together with those 2 resulting from eq.\ (\ref{TnnEoM}), 
we have $(d+1)$ such conditions. Hence, the number of the conditions 
is smaller than the number of the (gauge independent) degrees of 
freedom in $[K_{\mu\nu}]_+$ and $[\Ln\phi]_+$, i.e.\ these conditions 
can be in general fulfilled.

The situation is more interesting for the ``$-$'' side 
eq.\ (\ref{ThnEoM}). One could try to use the same 
argument as for the ``$-$'' side eq.\ (\ref{TnnEoM}) 
and expect to get additional consistency conditions 
for $[K_{\mu\nu}]_+$ and $[\Ln\phi]_+$. However, this time 
that argument does not work. The reason is that the 
difference between the ``$+$'' and the ``$-$'' sides 
projections (\ref{ThnEoM}) is closely related to the junction 
conditions (\ref{tensorJC}) and (\ref{scalarJC}). 
After a somewhat tedious calculation, it can be shown that 
such a difference is equivalent to the following condition 
on the brane sources:
\begin{equation}
D^\mu\tau_{\mu\nu} 
+ D^\mu\phi\,\big( h_{\mu\nu} \tau_\phi - \tau_{\mu\nu} \big)
=0\,.
\label{brane_conservation}
\end{equation}
This condition has the same simple form for all theories 
of the structure described by the Lagrangian (\ref{L}) 
of arbitrary higher order in derivatives. 
It does relate the brane sources (\ref{taus}) to the brane metric 
(the covariant derivative on the l.h.s.\ is covariant with respect 
to the induced brane metric $h_{\mu\nu}$). However, its character 
depends crucially on the brane localized contribution $\cL_B$
to the Lagrangian (\ref{L}). For a wide class of $\cL_B$ eq.\  
(\ref{brane_conservation}) is not a dynamical brane equation 
of motion, because it does not involve the second derivatives 
(in the brane directions) of the brane fields. 
It is rather a consistency condition on 
the source terms\footnote{
In the case without the dilaton, i.e.\ in the case of the Einstein-Lovelock 
theory of gravity, eq.\ (\ref{brane_conservation}) 
reduces to the requirement that the brane tensor source $\tau_{\mu\nu}$ 
must be covariantly conserved.
} 
$\tau_{\mu\nu}$ and $\tau_\phi$. Such consistency conditions are typical 
of gravity theories \cite{Wald}.

The situation changes for more complicated brane Lagrangians $\cL_B$. 
Specifically, if $\cL_B$ contains e.g.\ localized kinetic terms 
for the gravity \cite{k_gravity} or for the scalar \cite{k_scalar}, 
the condition (\ref{brane_conservation}) 
becomes a dynamical brane equation of motion involving the second 
derivatives (in the brane directions) of the brane fields $h_{\mu\nu}$ 
and $\phi$. Certainly even in such a case we do not 
get a full system of the brane gravitational equations of motion. 
The reason is obvious: the condition (\ref{brane_conservation}) is a 
$(d-1)$-dimensional vector of equations, while the  brane gravitational
equations should have a $(d-1)$-dimensional tensor character.

The condition (\ref{brane_conservation}) seems to be of the lowest order,
as all the higher order terms present in the bulk 
Lagrangian (\ref{L}) ``canceled-out'' in its derivation when we combined 
the higher order junction conditions (\ref{tensorJC}) and (\ref{scalarJC}) 
with the higher order bulk equation (\ref{ThnEoM}). However, 
the condition (\ref{brane_conservation}) may involve terms with more than 
two derivatives of the fields if appropriate interactions are present 
in the brane Lagrangian $\cL_B$.

Let us summarize the general case without $\Z2$ or any other 
symmetry imposed on the bulk Lagrangian (\ref{L}) or the bulk background 
solution. We have the following system of equations:
\begin{itemize}
\item[(i)]
Junction conditions (\ref{tensorJC}) and (\ref{scalarJC}), 
which determine (not always explicitly) the jumps in the values of the 
extrinsic curvature and the Lie derivative of the dilaton: 
$[K_{\mu\nu}]_\pm$ and $[\Ln\phi]_\pm$. 
\item[(ii)]
System of equations for $[K_{\mu\nu}]_+$ and $[\Ln\phi]_+$ 
given by the ``$+$'' side eq.\ (\ref{ThnEoM}) together with both 
eqs.\ (\ref{TnnEoM}). The number of these equations is smaller 
than the number of (gauge-independent) degrees of freedom 
in $[K_{\mu\nu}]_+$ and $[\Ln\phi]_+$, so that system of equations 
in general has solutions.
\item[(iii)]
Consistency condition (\ref{brane_conservation}) on the brane 
sources $\tau_{\mu\nu}$ and $\tau_\phi$.
\end{itemize}
The first two systems of equations, (i) and (ii), are not 
restrictive enough to obtain any effective brane equations of motion 
relating the dynamics of the brane fields to the brane sources (\ref{taus}). 
One can choose any brane fields, i.e.\ induced metric $h_{\mu\nu}$ 
and dilaton $\phi$, and any brane sources, $\tau_{\mu\nu}$ and 
$\tau_\phi$, satisfying the consistency condition (iii). 
In general, there exist values of $[K_{\mu\nu}]_{-(+)}$ and 
$[\Ln\phi]_{-(+)}$ which fulfill eqs.\ (i) and (ii) for any such 
choice. On the other hand, the consistency 
condition (iii) may have a character of a dynamical brane 
equation of motion. This depends crucially on the form of 
the brane localized interactions described by $\cL_B$.

%%%%%%%%%%%%%%%%%%%%%%%%%%%%%%%%
\subsection{ Construction with ${\Z2}$ symmetry}
\label{subsectionZ2}
%%%%%%%%%%%%%%%%%%%%%%%%%%%%%%%%

The bulk $\Z2$ symmetry is employed in many papers on brane models.  
We will show now that such a symmetry not only simplifies the 
calculations, but can also change qualitatively the problem of 
the existence of the effective brane equations of motion.

The bulk $\Z2$ symmetry relates the fields at the ``$+$'' and the ``$-$'' 
sides of the brane. We have e.g.
\begin{equation}
[K_{\mu\nu}]_+=-[K_{\mu\nu}]_-={\textstyle\frac12} \, [K_{\mu\nu}]_\pm
\,,
\qquad\qquad
[\Ln\phi]_+ =- [\Ln\phi]_- = {\textstyle\frac12} \, [\Ln\phi]_\pm
\,.
\label{Z2jc}
\end{equation}
In such a case, the ``$-$'' side eqs.\ (\ref{ThhEoM}), 
(\ref{WEoM})-(\ref{TnnEoM}) coincide with the ``$+$'' side ones. 
Moreover, the junction conditions (\ref{tensorJC}) and (\ref{scalarJC}) 
determine the extrinsic curvature $K_{\mu\nu}$ and the Lie derivative of the 
dilaton $\Ln\phi$ on both sides of the brane. The junction conditions 
are given by equations of order $(2N_{\rm max}-1)$, so in general
they cannot be solved explicitly for models of the higher order in derivatives. 
They can be solved analytically in some cases of highly symmetric 
configurations. Otherwise we have to solve them numerically.
Substituting the (explicit or not) solutions of the 
junction conditions, $[K_{\mu\nu}]_+(\tau_{\mu\nu},\tau_\phi)$ 
and $[\Ln \phi]_+(\tau_{\mu\nu},\tau_\phi)$, into eqs.\ (\ref{TnnEoM}) 
we obtain the following effective brane equation of motion:
\begin{eqnarray}
V(\phi)
-\sum_{N=1}^{N_{max}} \frac{\alpha_N}{2}\, \cT
\Bigg(
&&\hspace{-20pt}
\bigg[
    {\textstyle\frac12} R_{**}^{**} \oplus 2 (DD)_*^* \phi 
\oplus(-1) (D\phi)^2 
\nn[2pt]
&&\hspace{-10pt}
    \oplus(-1) \Big( 
[K^*_*]_+(\tau_{\mu\nu},\tau_\phi) \oplus(-1) 
[\Ln \phi]_+(\tau_{\mu\nu},\tau_\phi) \Big) ^2 \bigg]^N
\Bigg)
=0\,.
\label{EoM_Z2}
\end{eqnarray}
This is the first equation we found which has the character 
of an effective brane equation of motion even for simple 
brane Lagrangians $\cL_B$. It involves the brane sources (\ref{taus}), 
as well as the second derivatives (along the brane directions) 
of the brane metric, $R_{\mu\nu}^{\rho\sigma}$,
and the dilaton, i.e.\ $D^\nu D_\mu\phi$ and $(D\phi)^2$.
The covariant derivatives 
of the dilaton can be eliminated using the condition 
(\ref{brane_conservation}). This way we obtain one equation 
relating the dynamics of the brane metric $h_{\mu\nu}$ 
to the brane sources $\tau_{\mu\nu}$ and $\tau_\phi$. 
It is the only such equation which appears in our theory - if 
we assume nothing else but the $\Z2$ symmetry in the bulk.

One effective equation of motion in a $(d-1)$-dimensional 
brane gravity is not much. The induced brane metric $h_{\mu\nu}$ has 
$\,d(d-1)/2\,$ gauge independent degrees of freedom. Nevertheless, 
even one equation of motion can be very important if we restrict our 
attention to highly symmetric brane solutions. For example, 
any maximally symmetric space-time  (like de Sitter space-time used 
to describe inflation) is fully determined by 
just one parameter - the curvature scalar. Similarly, the 
cosmologically important Friedmann-Robertson-Walker space-time
depends on one function of time only - the cosmic scale factor,
for each sign of the brane spatial curvature.

Despite the fact that there is only one ``true'' effective brane equation
(\ref{EoM_Z2}), it is sometimes convenient to write down the full tensor 
Einstein-like equation of motion for the brane fields $h_{\mu\nu}$ and $\phi$. 
It follows straightforwardly from eq.\ (\ref{Thh_EE}) and is of the same 
structure, namely
\begin{eqnarray}
&& \hspace{-1.7cm}
{\textstyle\frac{(d-3)(2d-3)}{(d-1)(d-2)}}
h_{\rho\sigma} V
-{\textstyle\frac{d-3}{d-2}}\sum_{N=1}^{N_{\rm max}} 
\frac{\alpha_N}{2}\, h^\mu_\rho h^\nu_\sigma \, {\ocT}_{\mu\nu}
\bigg( \cM^N
\oplus{\textstyle\frac{2N}{d-3}}\cP\cM^{N-1}
\oplus\,8N(1-N) \,\cN \cM^{N-2}
%\hspace{-4pt}&&
\nn[2pt]
&&
    \oplus\, 2N 
\left[
h_*^* \frac{b_1 B_0 - b_0 B_1}{b_0 b_2 - b_1^2}
\oplus\frac{b_1 B_1 - b_2 B_0}{b_0 b_2 - b_1^2}
\right]
\cM^{N-1}
\bigg)
-{\textstyle\frac{d-3}{d-1}}\sum_{N=1}^{N_{\rm max}} 
\frac{\alpha_N}{2}\,h_{\rho\sigma} \,\cT\Big(\cM^N\Big)
=0
\, ,
%\hspace{12pt}&&
\label{Thh_EEZ2}
\end{eqnarray}
with the metric $h_{\mu\nu}$ and the dilaton $\phi$ taken on the 
brane and all (implicit) terms involving $[K_{\mu\nu}]_+$ and $[\Ln\phi]_+$ 
(with the bulk $\Z2$ symmetry it does not matter which side of the brane 
we choose) replaced with the appropriate solutions of the junction 
conditions (\ref{tensorJC}) and (\ref{scalarJC}). 
Obviously, the above effective brane equation does not 
constitute a closed system. 
The solution of the equations of motion for the bulk gravity 
has to be known to fully describe the gravity induced on the brane.
However, the total bulk dependence of the above equation is described 
by the projected Weyl tensor $E_{\mu\nu}$, given by (\ref{E}),
which enters (implicitly - via $\cP$ and $B_m$ 
defined in (\ref{cP}) and (\ref{Bm}), respectively) 
in a quite complicated and in general non-linear way.

Our effective brane equation (\ref{Thh_EEZ2}) 
 has one feature which is unusual for the standard 
equations of motion, but quite typical of the brane ones. 
Specifically, there are no terms linear in the brane energy-momentum 
tensor $\tau_{\mu\nu}$. The reason is as follows: 
The brane sources (in our model: $\tau_{\mu\nu}$ and $\tau_\phi$) 
appear explicitly in the junction conditions only, which determine 
the jumps in the values of the Lie derivatives 
(in our model: the extrinsic curvature, $K_{\mu\nu}$, and the Lie 
derivative of the dilaton, $\Ln\phi$)
of the brane fields 
(the induced metric tensor $h_{\mu\nu}$ and the dilaton field $\phi$, 
respectively). 
The left hand sides of the 
junction conditions (\ref{tensorJC}) and (\ref{scalarJC}) 
are given by polynomials of the order $(2N_{\rm max}-1)$ with only odd powers 
of the Lie derivatives. Although in general there can be several different 
solutions of these equations, we are interested only in those  
which vanish in the limit of vanishing sources (in the absence of sources the 
jumps in the values of the Lie derivatives have to vanish).  
Such solutions can be written as series in $\tau_{\mu\nu}$ and 
$\tau_\phi$ with vanishing constant terms. 
On the other hand, only even powers of the Lie derivatives 
($K_{\mu\nu}$ and $\Ln\phi$) are present in the (bulk)
equations of motion. Consequently, in our effective brane 
equation (\ref{Thh_EEZ2})
the terms  
involving the sources are at least quadratic (or bilinear) in 
$\tau_{\mu\nu}$ and $\tau_\phi$. 
Moreover, among the bilinear terms there is no term proportional 
to $\tau_\phi\tau_{\mu\nu}$. As we will show in the next section, 
such term is absent in the case of $N_{\rm max}=1$. It cannot 
appear for higher $N_{\rm max}$ as well, as the higher order 
corrections can change only those terms in the junction conditions 
solutions which are of higher order in the sources 
(we recall that one should consider only such solutions 
which have no constant terms when expanded in the sources).

In the standard Einstein equation the energy-momentum tensor 
appears linearly only. In order to have such a term in our 
effective brane equations of motion (\ref{EoM_Z2}) and (\ref{Thh_EEZ2}) 
we have to rewrite $\tau_{\mu\nu}$ as a sum of the energy-momentum 
tensor $\widetilde\tau_{\mu\nu}$ associated with the fields we are 
interested in (e.g.\ the Standard Model fields which  
are usually assumed to be localized on a brane) 
and some ``cosmological constant'' $\widetilde\lambda$ term:
$\tau_{\mu\nu}=\widetilde\tau_{\mu\nu}+h_{\mu\nu}\widetilde\lambda$. 
The result of such a redefinition will be discussed in more detail  
in the next section, which is devoted to the dilaton gravity 
with $N_{\rm max}=1$ and $N_{\rm max}=2$.

%%%%%%%%%%%%%%%%%%%%%%%%%%%%%%%%
%%%%%%%%%%%%%%%%%%%%%%%%%%%%%%%%
\section{Examples }
\label{section_examples}
%%%%%%%%%%%%%%%%%%%%%%%%%%%%%%%%
%%%%%%%%%%%%%%%%%%%%%%%%%%%%%%%%

The results presented in the previous sections are valid for 
models with corrections of arbitrarily high orders. 
In this section we write down those results 
explicitly\footnote{
Due to their complexity, explicit formulae leading to the $N_{\rm max}=2$ 
Einstein-like brane equation are moved to the appendix. 
}
for two simplest cases of $N_{\rm max}=1$ 
and  $N_{\rm max}=2$. Although models 
with higher order corrections are the main topic of this work, 
we nevertheless want to 
present the results also for the lowest order theory. 
There are two reasons. First: the complexity of the 
calculations grows rapidly with the order of
corrections.
Thus, it is more instructive to discuss the main features of our procedure 
and results in the simplest situation with $N_{\rm max}=1$. 
Second: we have obtained new results even for the lowest 
order theory. In our approach we treat the dilaton field 
on the same footing as the metric tensor, which has not been 
done before. The results on
the effective brane equations for the dilaton gravity 
obtained so far in the literature were not fully 
satisfactory\footnote{ 
For example, in ref.\ \cite{MaWa} not all the calculations 
were carried out in a fully covariant way. Terms containing the combination 
$a^\lambda\nabla_\lambda\phi$ were removed by a gauge choice. 
Moreover, the influence of the bulk scalar field on the brane 
dynamics was taken into account only in some approximation 
and its bulk behavior was not eliminated from the brane gravitational 
equations.
}.

In the present work we give only the general formulae 
for $N_{\rm max}=1$ and  $N_{\rm max}=2$. 
Presentation and discussion of some specific examples 
are postponed to a future publication.

%%%%%%%%%%%%%%%%%%%%%%%%%%%%%%%%
%%%%%%%%%%%%%%%%%%%%%%%%%%%%%%%%
\subsection{\boldmath $N_{\rm max}=1$ }
\label{subsection_N1}
%%%%%%%%%%%%%%%%%%%%%%%%%%%%%%%%
%%%%%%%%%%%%%%%%%%%%%%%%%%%%%%%%

Let us illustrate the main features of the construction of
the effective brane equations of motion by considering a 
simple example with $N_{\rm max}=1$. In this case
the ``$+$'' and the ``$-$'' sides limits of the bulk tensor 
equation of motion projected on the brane (\ref{ThhEoM}) 
have the form of
\begin{eqnarray}
&&\!\!\!\!\!\!\!\!\!\!\!\!
\bigg[ \Big\{
R_{\mu\nu} + (DD)_{\mu\nu}\phi 
- {\textstyle\frac12} h_{\mu\nu} \left[R +2 (DD)\phi - (D\phi)^2\right]
+ \alpha_1^{-1} V(\phi) h_{\mu\nu}  
\Big\}
+\,
\Big\{
2\,
(KK)_{\mu\nu} - K_{\mu\nu} (K-\Ln\phi)
%\quad
\nn[2pt]
&&\!\!\!\!\!\!\!\!
- {\textstyle\frac12}  h_{\mu\nu} \Big( 3(KK) - (K-\Ln\phi)^2  
- 2 a^\lambda \nabla_\lambda \phi \Big)
\Big\}
-\, \Big\{ 
\Ln K_{\mu\nu} - h_{\mu\nu} \big( (h\Ln K) - \Ln^2 \phi \big)
\Big\}
\bigg]_{-(+)}=0
\,,
\label{Thh_N1}
\end{eqnarray}
whereas the ``$+$'' and the ``$-$'' sides limits of the dilaton equation 
of motion (\ref{WEoM}) read
\begin{eqnarray}
&& \hspace{-1.9cm}
\bigg[
\Big\{R+2(DD)\phi-(D\phi)^2 - 
%{\textstyle\frac{2}{\alpha_1}} 
%\frac{2}{\alpha_1} 
2\alpha_1^{-1}
\big( V(\phi)
-V'(\phi) \big) \Big\}
+\Big\{3\,(KK)-(K-\Ln\phi)^2-2\,a^\lambda\nabla_\lambda\phi\Big\}
\nn[2pt]
&& \hspace{9cm}
-2\,\Big\{(h\Ln K)-\Ln^2\phi\Big\}
\bigg]_{-(+)}\!=0
\,.
\label{W_N1}
\end{eqnarray}
Both of these equations are genuine bulk equations of motion. 
They just determine the second Lie derivatives of the fields 
($\Ln K_{\mu\nu}$ and ${\Ln}^2\phi$ in the third curly bracket 
in each equation) in terms of the fields (the metric tensor $h_{\mu\nu}$ 
and the dilaton field $\phi$,  
together with their derivatives along the brane directions,  
in the first curly bracket) and their first Lie 
derivatives\footnote{
The last term in the second curly bracket is a mixture of the 
first Lie derivative of the dilaton with its derivatives along 
the brane directions. Hence, a part of that term should be moved to 
the first curly bracket. However, this subtlety does not change 
any further reasoning. Moreover, as was already mentioned, this 
slightly problematic term, $a^\lambda\nabla_\lambda\phi$, does not 
appear in our final results due to the appropriately designed 
derivation of the effective brane equations.
} 
($K_{\mu\nu}$ and $\Ln\phi$ in the second curly bracket). 
The values of all the quantities in the first two curly brackets 
in each equation are just the Cauchy boundary conditions for the 
corresponding $d$-dimensional second order differential equations. 
Without other equations or any additional assumptions about 
the bulk solution, these boundary conditions can be 
arbitrary. Thus eqs.\ (\ref{Thh_N1}) and (\ref{W_N1}) 
do not give any constraints on the brane fields $h_{\mu\nu}$ and $\phi$ 
in terms of the brane sources $\tau_{\mu\nu}$ and $\tau_\phi$. 
To get any effective brane equations 
of motion we have to analyze the junction conditions 
and the bulk tensor equation of motion with at least one index contracted 
with that of the vector $n^\mu$ normal to the brane.

The junction conditions (\ref{tensorJC}) and (\ref{scalarJC}) 
are very simple in the case of $N_{\rm max}=1$. 
Solving them we can express the jumps in the values of 
the extrinsic curvature, $K_{\mu\nu}$, and the Lie derivative of the 
scalar field, $\Ln\phi$, in terms of the brane sources, 
$\tau_{\mu\nu}$ and $\tau_\phi$, as
\begin{eqnarray}
\left[K_{\mu\nu}\right]_\pm 
= \alpha_1^{-1} \big(h_{\mu\nu}\tau_\phi-\tau_{\mu\nu}\big)
\,,
\label{JC_N1solved1}
\\[2pt]
\left[\Ln\phi\right]_\pm
= \alpha_1^{-1} \big((d-2)\tau_\phi-\tau\big)
\,.
\label{JC_N1solved2}
\end{eqnarray}
Subsequently, we consider two vector eqs.\ (\ref{ThnEoM})
- one for each side of the brane. The difference of those two equations 
together with the junction conditions (\ref{JC_N1solved1}) and 
(\ref{JC_N1solved2}) gives the consistency condition 
(\ref{brane_conservation}) for the sources, namely
\begin{equation}
D^\mu\tau_{\mu\nu} 
+ D^\mu\phi \, \big( h_{\mu\nu} \tau_\phi - \tau_{\mu\nu} \big)
=0\,.
\label{brane_conservationN1}
\end{equation}
As the second combination of the ``$-$'' and the ``$+$'' side eqs.\ 
(\ref{ThnEoM}) we take the ``$+$'' one. 
Hence, we obtain the following condition for 
the ``$+$'' side quantities:
\begin{equation}
D^\mu \big[K_{\mu\nu}\big]_+ 
- D^\mu\phi \big[K_{\mu\nu}\big]_+
-D_\nu \big[K\big]_+ + D_\nu\big[\Ln\phi\big]_+=0
\,.
\label{Thn_N1+-}
\end{equation}
Finally, we have to take into account two scalar eqs.\ 
(\ref{TnnEoM}) with $N_{\rm max}=1$.  
The brane curvature $R$ and the dilaton $\phi$, together with 
the first and the second (covariant with respect to the brane metric $h_{\mu\nu}$) 
derivatives of the latter all are continuous at the brane. Thus their 
contributions cancel in the difference of the ``$-$'' and the ``$+$'' 
side eqs.\ (\ref{TnnEoM}). Such a difference reduces to the equality 
$0= \left[ ( K - \Ln\phi )^2\right]_\pm - \left[ (KK)\right]_\pm$,
which, after employing the junction conditions (\ref{JC_N1solved1}) 
and (\ref{JC_N1solved2}), can be rewritten as
\begin{equation}
\tau^{\mu\nu}\left[K_{\mu\nu}\right]_+
-\tau_\phi\left[\Ln\phi\right]_+
+{\textstyle\frac12} \alpha_1^{-1} \big((\tau\tau)-2\tau\tau_\phi
+(d-2)\tau_\phi^2\big)
=0
\, ,
\label{Thn_N1+}
\end{equation}
where $(\tau\tau)\equiv\tau^{\mu\nu}\tau_{\mu\nu}$. 
The ``$+$'' side eq.\ (\ref{TnnEoM}) reads
\begin{equation}
 R + 2 (DD)\phi - (D\phi)^2 - 2\alpha_1^{-1}V(\phi)
=
 \left( \left[K\right]_+ - \left[\Ln\phi\right]_+ \right)^2 
- \left[ (KK) \right]_+
\, .
\label{Tnn_N1+}
\end{equation}

The last six eqs.\ (\ref{JC_N1solved1})-(\ref{Tnn_N1+}) are the only 
equations which can yield the effective brane equations of motion 
in a general case.  
The junction conditions (\ref{JC_N1solved1}) and (\ref{JC_N1solved2}) 
determine the jumps in the values 
of the Lie derivatives of the metric tensor and the dilaton field: 
$[K_{\mu\nu}]_\pm$ and $[\Ln\phi]_\pm$, respectively. 
These quantities do not appear in any of the remaining four equations. 
Equation (\ref{brane_conservationN1}) is a consistency condition on the 
brane sources  $\tau_{\mu\nu}$ and $\tau_\phi$ - corresponding to the 
covariant conservation of the energy momentum tensor in the standard 
theory of gravity. The remaining equations establish conditions for 
the directional limits of the values of the first derivatives (normal 
to the brane) of the brane fields - evaluated at one of the 
``sides of the brane'', which we chose to be the ``$+$'' side. 
Specifically, eq. (\ref{Thn_N1+}) relates 
$[K_{\mu\nu}]_+$ and $[\Ln\phi]_+$ 
to the brane sources $\tau_{\mu\nu}$ and $\tau_\phi$, 
while eqs.\ (\ref{Thn_N1+-}) and (\ref{Tnn_N1+}) 
relate them to the brane fields $h_{\mu\nu}$ and $\phi$. Equations 
(\ref{Thn_N1+-})-(\ref{Tnn_N1+}) provide us with $(d+1)$ relations, 
i.e.\ less than 
the number of the (gauge-independent) degrees of freedom in 
 $[K_{\mu\nu}]_+$ and $[\Ln\phi]_+$. Thus, in general, eqs.\  
(\ref{Thn_N1+-})-(\ref{Tnn_N1+}) can be solved for arbitrary 
brane fields and sources. 
The junction conditions (\ref{JC_N1solved1}) and (\ref{JC_N1solved2}) 
do not change this situation, as without any assumptions on the 
bulk solution (as e.g.\ the already mentioned and usually employed 
bulk $\Z2$ symmetry) the quantities $[K_{\mu\nu}]_\pm$, $[\Ln\phi]_\pm$, 
$[K_{\mu\nu}]_+$ and $[\Ln\phi]_+$ are all independent.

The only equation which involves the brane quantities exclusively is 
given by the formula (\ref{brane_conservationN1}). However, it is usually 
considered as a consistency condition on the brane sources 
and not as a dynamical equation of motion. As we pointed out in 
the previous section, it can yield a dynamical equation if the brane 
Lagrangian $\cL_B$ is complicated enough, e.g.\ when it involves 
brane localized kinetic terms for the bulk fields.

Let us now check the implications of some usually assumed features 
of the bulk solution. The by far most popular assumption of this kind
is the already discussed bulk $\Z2$ symmetry with the fixed point at the 
brane position. With this symmetry it is enough to consider eqs.\ 
(\ref{ThnEoM}) and (\ref{TnnEoM}) on one ``side of the brane'' only. 
The corresponding equations on the other side of the brane are fulfilled 
automatically. Hence, in the case of $N_{\rm max }=1$ we are left with 
eqs.\ (\ref{JC_N1solved1})-(\ref{brane_conservationN1}) and 
(\ref{Tnn_N1+}). Employing
$[K_{\mu\nu}]_\pm=2[K_{\mu\nu}]_+$ and $[\Ln\phi]_\pm=2[\Ln\phi]_+$ 
relations, which are due to the bulk $\Z2$ symmetry, together with the 
junction conditions (\ref{JC_N1solved1}) and (\ref{JC_N1solved2}), 
we can rewrite eq.\ (\ref{Tnn_N1+}) in the following form:
\begin{equation}
 R + 2 (DD)\phi - (D\phi)^2 - 2\alpha_1^{-1}V(\phi)
=
{\textstyle\frac14}\alpha_1^{-2}
\left[ \, -(\tau\tau) + 2\, \tau\tau_\phi - (d-2)\, \tau_\phi^2 \, \right]
.
\label{braneEoM}
\end{equation}
It determines the brane  
curvature scalar $R$ in terms of the brane sources 
$\tau_{\mu\nu}$ and $\tau_\phi$ (the dilaton 
field derivative can be obtained from the consistency condition 
(\ref{brane_conservationN1})).

We shall now construct explicitly the brane tensor equation
of motion for $N_{\rm max}=1$ - following the procedure 
for arbitrary $N_{\rm max}$, described at the end of subsection 
\ref{subsection_general_case}.  
We start with eliminating the second derivatives normal to the brane,
$\Ln K_{\mu\nu}$ and $\Ln^2\phi$,  
from the bulk tensor equation (\ref{Thh_N1}) and 
the bulk dilaton equation of motion (\ref{W_N1}). 
The parameters $b_m$ and $B_m$, defined 
in (\ref{bm}) and (\ref{Bm}), can be easily calculated: 
\begin{eqnarray}
b_m
=
&&\hspace{-18pt}
\alpha_1\,{\textstyle\frac{(d-1)!}{(d-1-m)!}}
\,,
\label{bmN1}
\\[2pt]
B_m 
=
&&\hspace{-18pt}
\alpha_1
\Big\{
{\textstyle\frac12}(d-3)^m\left[R-K^2+(KK)\right]
+(d-2)^m\big[(DD)\phi+K\Ln\phi\big]
\nn[2pt]
&&
-{\textstyle\frac12}(d-1)^m\left[(D\phi)^2+(\Ln\phi)^2\right]
\Big\}
-(d-1)^m V
+(1-m) V'
\,.
\label{BmN1}
\end{eqnarray}
Employing the above explicit formulae on the parameters $b_m$ and $B_m$,
and the definitions (\ref{cM}) and (\ref{cP}),
with $K_{\mu\nu}$ and $\Ln\phi$ given by their limits 
$[K_{\mu\nu}]_+$ and $[\Ln\phi]_+$ obtained from
the junction conditions (\ref{JC_N1solved1}) and (\ref{JC_N1solved2})
supplemented by the relations (\ref{Z2jc}) due to the bulk $\Z2$ symmetry,
the brane tensor equation (\ref{Thh_EEZ2}) reduces to
\begin{eqnarray}
R_{\mu\nu} - {\textstyle\frac12} h_{\mu\nu} R
=
&&\hspace{-18pt}
- {\textstyle\frac{d-3}{d-2}} \big[ (DD)_{\mu\nu}\phi -  h_{\mu\nu} (DD)\phi\big] 
- {\textstyle\frac{d-3}{d-1}} h_{\mu\nu} 
\big[ \textstyle\frac12(D\phi)^2 + \alpha_1^{-1} V(\phi) \big]
- E_{\mu\nu} 
\nn[2pt]
&&\hspace{-18pt}
%:
 + {\textstyle\frac14}\alpha_1^{-2} \left[  
{\textstyle\frac{1}{d-2}} \tau \tau_{\mu\nu} - (\tau\tau)_{\mu\nu} 
+  h_{\mu\nu} \left( {\textstyle\frac12} (\tau\tau)
- {\textstyle\frac{1}{(d-1)(d-2)}} \tau^2
\right.\right.
\nn[2pt]
&& \hspace{172pt}
\left.\left.
- {\textstyle\frac{d-3}{d-1}} \tau \tau_\phi 
+ {\textstyle\frac{(d-2)(d-3)}{2(d-1)}} \tau_\phi^2 \right) \right] 
.
\label{EeN1Z2}
\end{eqnarray}
This equations is truly of the form of the $(d-1)$-dimensional 
Einstein equation with some corrections (which was not apparent 
for the general $N_{\rm max}$ formula (\ref{Thh_EEZ2})). 
Three specific types of contributions can be discerned on its r.h.s. 
There are terms with the explicit 
$\phi$-dependence, typical of the gravity theories with 
scalar fields. Moreover, the tensor $E_{\mu\nu}$ represents the bulk 
influence on the dynamics at the brane. Finally, the last square 
bracket contains the contributions from the brane sources 
$\tau_{\mu\nu}$ and $\tau_\phi$. 
These contributions are quadratic in the brane energy-momentum 
tensor $\tau_{\mu\nu}$ (which is typical of the brane models) and its 
dilaton counterpart $\tau_\phi$. 
In order to have terms linear in some energy-momentum tensor 
(as is the case in the standard Einstein gravity) we have to rewrite 
$\tau_{\mu\nu}$ as the already mentioned sum 
of the energy-momentum tensor $\widetilde\tau_{\mu\nu}$ associated with 
the fields we are interested in and some 
``cosmological constant'' $\widetilde\lambda$ term:
$\tau_{\mu\nu}=\widetilde\tau_{\mu\nu}+h_{\mu\nu}\widetilde\lambda$. 
With such a decomposition we get the Einstein-like brane equation 
of motion as
\begin{eqnarray}
R_{\mu\nu} - {\textstyle\frac12} h_{\mu\nu} R
=
&&\hspace{-18pt}
8\pi\widetilde G \, \widetilde\tau_{\mu\nu} 
- {\textstyle\frac{d-3}{d-2}} \big[ (DD)_{\mu\nu}\phi -  h_{\mu\nu} (DD)\phi\big] 
- {\textstyle\frac{d-3}{d-1}} h_{\mu\nu} 
\big[ \textstyle\frac12(D\phi)^2 + \alpha_1^{-1} V(\phi) \big]
- E_{\mu\nu} 
\nn[2pt]
&&\hspace{-18pt}
 + {\textstyle\frac14}\alpha_1^{-2} \left[  
{\textstyle\frac{1}{d-2}} \, \widetilde\tau\widetilde\tau_{\mu\nu}
- (\widetilde\tau\widetilde\tau)_{\mu\nu} 
+ {\textstyle\frac12} h_{\mu\nu} \, (\widetilde\tau\widetilde\tau)
- {\textstyle\frac{1}{(d-1)(d-2)}} h_{\mu\nu}\, {\widetilde\tau}^2
+ \left( {\textstyle\frac{d-3}{d-2}} \, \widetilde\lambda 
- {\textstyle\frac{d-3}{d-1}} \, \tau_\phi \right) h_{\mu\nu} \, \widetilde\tau 
\right.
\nn[2pt]
&& \hspace{32pt}
\left.
+ \left({\textstyle\frac{d-3}{2}}\,\widetilde\lambda^2 
- (d-3)\,\widetilde\lambda\,\tau_\phi 
+ {\textstyle\frac{(d-2)(d-3)}{2(d-1)}} \, \tau_\phi^2\right) h_{\mu\nu}
\right]
,
\label{EeN1Z2redefined}
\end{eqnarray}
where we introduced 
\begin{equation}
\widetilde G \equiv \frac{-(d-3)\,\widetilde\lambda}{32(d-2)\pi\,\alpha_1^2}
\,,
\label{newton}
\end{equation}
which can be interpreted as the effective brane Newton's constant. 
The contributions to eq.\ (\ref{EeN1Z2redefined}) which are proportional 
to the brane metric, $h_{\mu\nu}$, and depend neither on the  
scalar field $\phi$, nor on the tensor $\widetilde\tau_{\mu\nu}$, 
should be interpreted as the effective brane cosmological constant:
\begin{equation}
\widetilde\Lambda
=
{\textstyle\frac{d-3}{d-1}}\,\alpha_1^{-1}V|_{\phi=0}
- {\textstyle\frac14}\,\alpha_1^{-2} \left( 
{\textstyle\frac{d-3}{2}}\,\widetilde\lambda^2 
- (d-3)\,\widetilde\lambda\,\tau_\phi|_{\phi=0} 
+ {\textstyle\frac{(d-2)(d-3)}{2(d-1)}} \, \tau_\phi^2|_{\phi=0}
\right)
.
\label{cosmoc}
\end{equation}
It depends on the brane Lagrangian $\cL_B$ (via $\tau_\phi$)
and on the way in which we divide the brane energy-momentum 
tensor $\tau_{\mu\nu}$ into its ``standard'' part $\widetilde\tau_{\mu\nu}$ 
and the ``cosmological constant'' term $h_{\mu\nu}\widetilde\lambda$. 
In addition, a part of $E_{\mu\nu}$ proportional to $h_{\mu\nu}$
can be also treated as a contribution to the effective brane cosmological 
constant.

The effective brane equation of motion (\ref{EeN1Z2redefined}) has the 
same tensor character as the standard Einstein equation.  
This certainly does not mean that we found another bulk-independent 
effective brane equation of motion - in addition to (\ref{braneEoM}). 
It is just a convenient way to parameterize the bulk influence 
by a single geometric quantity: the projection of the bulk Weyl tensor on 
the brane, $E_{\mu\nu}$. Observe that there is no additional bulk 
influence due to the presence of the dilaton 
field\footnote
{This differs from the results previously presented in the literature.
For example, it is claimed in \cite{MeBa} that in the 
effective brane equations of motion it is easier 
to remove the dependence on the projected bulk Weyl tensor than 
the dependence on the bulk dilaton field. Our analysis clearly indicates 
the opposite. 
}.

%%%%%%%%%%%%%%%%%%%%%%%%%%%%%%%%
%%%%%%%%%%%%%%%%%%%%%%%%%%%%%%%%
\subsection{\boldmath $N_{\rm max}=2$ }
\label{subsection_N2}
%%%%%%%%%%%%%%%%%%%%%%%%%%%%%%%%
%%%%%%%%%%%%%%%%%%%%%%%%%%%%%%%%

The formulae for arbitrary order of corrections given 
in sections \ref{section_JC} and \ref{section_effectiveEoM} 
can be employed to obtain more explicit equations for 
any given $N_{\rm max}$. However, the complexity of the 
resulting expressions grows rapidly with $N_{\rm max}$. 
The effective brane equations of motion become quite 
intricate already for the $N_{\rm max}=2$ case. 
Nevertheless, they can be still obtained 
even in the most general case, i.e.\ without any additional assumptions 
on the bulk background. We impose only the usual 
$\Z2$ symmetry. 
As was already underlined, the $N_{\rm max}=2$ case is equivalent
to the appropriate subset of higher order interaction terms of 
the effective action derived from string theories.

In the $N_{\rm max}=2$ case the junction conditions 
(\ref{tensorJC}) and (\ref{scalarJC}) take the following form:
\begin{eqnarray}
\tau_{\mu\nu}
=
&&\hspace{-18pt}
2\Big[
\alpha_1\big[h_{\mu\nu}(K-\Ln\phi)-K_{\mu\nu}\big]
\nn[2pt]
&&\hspace{-12pt}
+\,2\,\alpha_2
\Big\{
\big[h_{\mu\nu}(K-\Ln\phi)-K_{\mu\nu}\big]
\left[R-K^2+(KK)+2\,(DD)\phi+2\,K\Ln\phi-(D\phi)^2-(\Ln\phi)^2\right]
\nn[2pt]
&&\hspace{24pt}
-\,2\,h_{\mu\nu}\,K_{\rho\sigma}
\big[R^{\rho\sigma}+(KK)^{\rho\sigma}+(DD)^{\rho\sigma}\phi\big]
-2\,(K-\Ln\phi)\big[R_{\mu\nu}+(DD)_{\mu\nu}\phi+(KK)_{\mu\nu}\big]
\nn[2pt]
&&\hspace{24pt}
+\,2\,K_{\mu\rho}\big[R^{\rho}_{\nu}+(KK)^{\rho}_{\nu}+(DD)^{\rho}_{\nu}\phi\big]
+2\,K_{\nu\rho}\big[R^{\rho}_{\mu}+(KK)^{\rho}_{\mu}+(DD)^{\rho}_{\mu}\phi\big]
\nn[2pt]
&&\hspace{24pt}
+\,2\,K^{\rho\sigma}\big[R_{\mu\rho\nu\sigma}-K_{\mu\rho}K_{\nu\sigma}\big]
+{\textstyle\frac23}\big[h_{\mu\nu}(K-\Ln\phi)^3+2(KKK)\big]
\Big\}
\Big]_+\,,
\label{tensorJCN2}
%\end{eqnarray}
%
%
%
%
%\begin{eqnarray}
\\
\tau_\phi
=
&&\hspace{-18pt}
2\Big[
\alpha_1 \big(K-\Ln\phi\big)
%\nn[2pt]
%&&
+2\,\alpha_2
\Big\{\!
(K-\Ln\phi)
\big[R-K^2+(KK)+2\,(DD)\phi+2\,K\Ln\phi-(D\phi)^2-(\Ln\phi)^2\big]
\!\!\nn[2pt]
&&\hspace{107pt}
+\,{\textstyle\frac23}\,(K-\Ln\phi)^3
-2K_{\mu\nu}\big[R^{\mu\nu}+{\textstyle\frac13}(KK)^{\mu\nu}
+(DD)^{\mu\nu}\phi\big]
\Big\}
\Big]_+
\,,
\label{scalarJCN2}
\end{eqnarray}
where $(KKK)$ denotes the trace of the third power 
of the extrinsic curvature: 
$K_\rho^\sigma K_\sigma^\lambda K_\lambda^\rho$. 
There are two new features as compared to the lowest order case. 
First: the junction conditions (\ref{tensorJCN2})
and (\ref{scalarJCN2}) are no longer linear 
in $\left[K_{\mu\nu}\right]_+$ and $\left[\Ln\phi\right]_+$. 
They are now third order equations for these quantities 
and their tensor structure is much more complicated.  
Thus, obtaining an explicit result is considerably more difficult. 
Second: solving these junction conditions (explicitly or not) 
yields the jumps in the values of $K_{\mu\nu}$ and $\Ln\phi$ 
as functions not only of the brane sources $\tau_{\mu\nu}$ 
and $\tau_\phi$,  but also of the brane curvature $R$ and the dilaton $\phi$.

The simplest form of the effective brane equation of motion 
is given by the scalar eq.\ (\ref{TnnEoM}). 
In the model with the bulk $\Z2$ symmetry 
and $N_{\rm max}=2$ it reads:
\begin{eqnarray}
V
&&\hspace{-18pt}
-{\textstyle\frac12}\,\alpha_1
\left[
R - K^2 + (KK) +2\,(DD)\phi + 2\,K\Ln\phi - (D\phi)^2 - (\Ln\phi)^2
\right]
\nn[2pt]
&&\hspace{-18pt}
-{\textstyle\frac12}\,\alpha_2
\Big\{
\left[
R - K^2 + (KK) +2\,(DD)\phi + 2\,K\Ln\phi - (D\phi)^2 - (\Ln\phi)^2
\right]^2
\nn[2pt]
&&\hspace{21pt}
-\,4\big[
R_{\mu\nu} - KK_{\mu\nu} + (KK)_{\mu\nu} + (DD)_{\mu\nu}\phi
+ K_{\mu\nu}\Ln\phi 
\big]\cdot
\nn[2pt]
&&\hspace{35pt}
\cdot\big[
R^{\mu\nu} - KK^{\mu\nu} + (KK)^{\mu\nu} + (DD)^{\mu\nu}\phi
+ K^{\mu\nu}\Ln\phi 
\big]
\nn[2pt]
&&\hspace{21pt}
+\,\big[
R_{\mu\nu\rho\sigma} - K_{\mu\rho}K_{\nu\sigma} + K_{\mu\sigma}K_{\nu\rho}
\big]\big[
R^{\mu\nu\rho\sigma} - K^{\mu\rho}K^{\nu\sigma} + K^{\mu\sigma}K^{\nu\rho}
\big]
\Big\}=0
\,,
\label{TnnEoMN2}
\end{eqnarray}
with $K_{\mu\nu}$ (and all its contractions) and $\Ln\phi$ 
replaced by their ``next to the brane'' values $[K_{\mu\nu}]_+$  
and $[\Ln\phi]_+$, given by the solutions of the junction 
conditions (\ref{tensorJCN2}) and (\ref{scalarJCN2}).
This is the only bulk-independent brane equation of motion in addition to  
the consistency condition (\ref{brane_conservation}) on the brane sources
$\tau_{\mu\nu}$ and $\tau_\phi$. 
It is also possible to derive the Einstein-like effective equation of motion,
as discussed at the end of subsection \ref{subsection_general_case}. 
However, the result is 
truly complicated and we postpone presenting the appropriate formulae 
to the appendix. Those formulae indicate that employing the brane 
Einstein-like effective equation for a general case is rather 
problematic. However, it can be significantly simplified if we restrict 
our considerations to the situations with sufficient symmetries. 
This is quite typical of all theories of gravity. 
Usually only solutions with some specific 
symmetry properties are looked for. An alternative way to investigate 
some highly symmetric solutions is to use the simplest form 
of the effective brane equation (\ref{TnnEoMN2}) instead 
of the Einstein-like form discussed in the appendix.

%%%%%%%%%%%%%%%%%%%%%%%%%
\section{ Conclusions }
\label{section_conclusions}
%%%%%%%%%%%%%%%%%%%%%%%%%  

The starting point of the present analysis has been given by 
the $d$-dimensional higher order dilaton gravity constructed 
previously \cite{KoOl} as a generalization of the Einstein-Lovelock 
theory and remaining in the close relation to the effective action 
in string theories if restricted to the gravity and the dilaton field.
It was supplemented by a co-dimension 1 brane with general 
brane localized interactions $\cL_B$ included into the Lagrangian (\ref{L}). 
The effective brane equations of 
motion for such a theory were constructed 
and discussed. All 
calculations were 
performed in the covariant approach.

In order to obtain the effective brane equations of motion one has to start 
from the full bulk equations of motion (\ref{PtensorEoM}), (\ref{PscalarEoM}) 
and attempt to eliminate all quantities evaluated away from the brane 
position. In general this is not possible and the effective brane equations 
do not form a closed system. The dynamics of the brane fields 
$h_{\mu\nu}$ (the induced brane metric tensor) and $\phi$ (the dilaton 
field) depends usually not only on the brane sources $\tau_{\mu\nu}$ and 
$\tau_\phi$ defined in (\ref{taus}), but also on the bulk gravity solution. 

On the basis of the full bulk equations of motion
derived from the Lagrangian density defining our model, 
we can obtain three types 
of equations involving fields determined either on or 
infinitesimally close to the brane.
These equations represent the junction conditions 
and two directional limits of the bulk equations established
when the brane is approached 
from the ``$+$'' and the ``$-$'' sides, respectively.
The junction conditions (\ref{tensorJC}) and (\ref{scalarJC})
relate the brane fields, $h_{\mu\nu}$ and $\phi$, 
and the sources, $\tau_{\mu\nu}$ and $\tau_\phi$,
to the across-the-brane jumps in the values of the extrinsic curvature 
and the Lie derivative (along the vector field $n^\mu$ orthonormal 
to the brane) of the scalar field: 
$\left[ K_{\mu\nu} \right]_\pm$ and $\left[ \Ln\phi \right]_\pm$,
respectively. The directional limits 
of the bulk equations (given by eqs.\ (\ref{ThhEoM}), 
(\ref{WEoM})-(\ref{TnnEoM}))
involve the brane fields ($h_{\mu\nu}$ and $\phi$),
together with their first ($K_{\mu\nu}$ and $\Ln\phi$) 
and second ($\Ln K_{\mu\nu}$ and $\Ln^2\phi$) Lie derivatives, 
but do not include the 
brane sources $\tau_{\mu\nu}$ and $\tau_\phi$. 
An important point should be underlined: the number of  
the $d$-dimensional bulk gravitational equations of motion 
is bigger than the number of equations necessary for a  
$(d-1)$-dimensional brane gravity. These 
extra equations (corresponding to   
the consistency conditions for a Cauchy problem with 
the boundary conditions defined infinitesimally close 
to the brane) play a crucial role, as they do 
not depend on the second Lie derivatives of the fields,
i.e.\ $\Ln K_{\mu\nu}$ and $\Ln^2\phi$.

Genuine effective brane equations of motion should relate 
the values of the fields, $h_{\mu\nu}$ and $\phi$, 
evaluated on the brane, 
to the brane sources, $\tau_{\mu\nu}$ and $\tau_\phi$. 
In addition, they should not 
depend on the bulk configuration. The crucial
question is: How many such brane equations can be 
obtained by combining information from all bulk 
equations mentioned in the previous paragraph? The answer 
depends strongly on the symmetries we assume for 
the bulk theory and for the bulk and/or brane solutions. 
The minimal number of the effective brane equations is obtained 
when no such symmetries are assumed. The only effective brane 
equation in such a case is given by formula (\ref{brane_conservation}). 
However, this equation is usually treated not as a dynamical 
brane equation of motion, but rather as a consistency 
condition on the brane sources (\ref{taus}). In the pure gravity case 
(i.e.\ without the dilaton) it reduces simply to the covariant 
conservation of the brane localized energy-momentum tensor. 
However, there are models in which this ``consistency'' 
condition yields a dynamical equation of motion. This is the case 
when e.g.\ brane localized kinetic terms are
present in the brane localized Lagrangian $\cL_B$.

The number of the effective brane equations of motion increases 
when we restrict the model by imposing some symmetries. 
The bulk $\Z2$ symmetry with the fixed point coinciding with 
the brane position is particularly frequently employed. In such a case, 
an additional effective brane equation of motion (\ref{EoM_Z2}) appears. 
It is obtained from the bulk tensor equation of motion if 
both indices are contracted with the vector field $n^\mu$ normal to the
brane - after applying the appropriate junction conditions. 
This is the only effective brane equation of motion 
in models for which the ``consistency'' condition  
(\ref{brane_conservation}) is not dynamical. 
The importance of this equation depends on the class 
of solutions we are interested in. For example, 
it is all we need when considering maximally symmetric 
brane solutions.

In many models the number of the effective brane equations 
of motion is (much) smaller than the number of 
independent components of the Einstein equation
in a $(d-1)$-dimensional space-time. Nevertheless, 
it is useful to derive an Einstein-like (tensor) brane equation. 
Such an equation for the higher (arbitrary) order dilaton gravity with 
the bulk $\Z2$ symmetry is given by eq.\ (\ref{Thh_EEZ2}). 
The entire dependence of the brane dynamics on the bulk 
gravity solution is encoded in the Weyl tensor projected on 
the brane, $E_{\mu\nu}$, which appears in eq.\ (\ref{Thh_EEZ2})
implicitly via the parameters defined by formulae
(\ref{Bm}) and (\ref{cP}). It should be stressed 
that, contrary to some previous claims, those effective 
brane gravitational equations do not depend on the 
bulk scalar solution.

Our general results obtained for corrections up to order $2N_{\rm max}$
(arbitrary, as long as it is not higher than the space-time dimensionality)
in derivatives are presented in a very compact notation based on
the generalizations (\ref{cT}) and (\ref{cTbar}) of 
the trace operation. Those results are rewritten explicitly 
in the conventional notation for the two simplest cases 
of $N_{\rm max}=1$ and $N_{\rm max}=2$. 
Although the analysis of the higher order theories is the main topic 
of our work, two reasons motivated us to address also the lowest 
order theory (i.e.\ with $N_{\rm max}=1$),
which has been already considered by other authors. 
First, it is useful as the simplest illustration of our general, 
non-trivial procedure. 
Second, we improved the analysis presented so far in the literature
even for this lowest order theory.
The resulting explicit formulae for the effective brane Newton's and 
cosmological constants are given by 
eqs.\  (\ref{newton}) and (\ref{cosmoc}), respectively.

The case of $N_{\rm max}=2$ is the 
Einstein-Gauss-Bonnet gravity interacting with a scalar field 
(which is also self-interacting) via terms with up to 
four derivatives. Similarly to the $N_{\rm max}=1$ case, 
the effective brane equations of motion do not 
involve quantities dependent on the bulk scalar solution. The total
bulk influence enters again through the projected Weyl tensor 
$E_{\mu\nu}$. The effective brane equations of motion for 
such a theory have not been presented before. They are 
quite lengthy when the generalized traces are explicitly 
calculated (all necessary formulae are collected in the 
appendix). Although the effective brane equations are rather 
complicated for a general case, they simplify substantially if
highly symmetric branes are considered. 
Applications of the derived equations for such symmetric models 
are postponed to a future publication.

%%%%%%%%%%%%%%%%%%%%%%%%%
\section*{Acknowledgments}
%%%%%%%%%%%%%%%%%%%%%%%%%  

This work was partially supported by the EC 6th 
Framework Projects 
MTKD-CT-2005-029466 ``Particle Physics and Cosmology:
the Interface'' and MRTN-CT-2006-035863 
``The Origin of Our Universe: Seeking Links between 
Fundamental Physics and Cosmology''.  
D.K.\ acknowledges partial support from the Polish 
MNiSW grant N202 175335. 
M.O.\ would like to thank for the hospitality experienced 
at Max Planck Institute for Physics in Munich, where 
part of this work has been prepared.

%%%%%%%%%%%%%%%%%%%%%%%%%
\section*{Appendix}
\renewcommand{\theequation}{A.\arabic{equation}}
\setcounter{equation}{0}
%%%%%%%%%%%%%%%%%%%%%%%%%  

In the appendix we collect the formulae appearing in the 
Einstein-like effective brane equation (\ref{Thh_EEZ2}) 
for the case of $N_{\rm max}=2$. This equation, 
after writing down explicitly the sum over $N$, takes the form of
\begin{eqnarray}
&&\hspace{-41pt}
{\textstyle\frac{(d-3)(2d-3)}{(d-1)(d-2)}}
h_{\rho\sigma} V
-{\textstyle\frac12}\,\alpha_1
\bigg\{
{\textstyle\frac{d-3}{d-2}}
h^\mu_\rho h^\nu_\sigma \, {\ocT}_{\mu\nu}\left( \cM \right)
+{\textstyle\frac{2}{d-3}}
h^\mu_\rho h^\nu_\sigma \, {\ocT}_{\mu\nu}\left( \cP \right)
+{\textstyle\frac{d-3}{d-1}}
h_{\rho\sigma}\,\cT\left(\cM\right)
\nn[2pt]
&&\hspace{120pt}
+\,2\,(d-3)
\frac{b_1 B_0 - b_0 B_1}{b_0 b_2 - b_1^2}h_{\rho\sigma}
+{\textstyle\frac{2}{d-2}}
\frac{b_1 B_1 - b_2 B_0}{b_0 b_2 - b_1^2}h_{\rho\sigma}
\bigg\}
\nn[2pt]
&&\hspace{-25pt}
-\,{\textstyle\frac12}\,\alpha_2
\bigg\{
{\textstyle\frac{d-3}{d-2}}
h^\mu_\rho h^\nu_\sigma \, {\ocT}_{\mu\nu}\left( \cM^2 \right)
+{\textstyle\frac{4}{d-3}}
h^\mu_\rho h^\nu_\sigma \, {\ocT}_{\mu\nu}\left( \cP\cM \right)
+{\textstyle\frac{d-3}{d-1}}
h_{\rho\sigma}\,\cT\left(\cM^2\right)
-{\textstyle\frac{16(d-3)}{d-2}}
h^\mu_\rho h^\nu_\sigma \, {\ocT}_{\mu\nu}\left( \cN \right)
\nn[2pt]
&&\hspace{19pt}
+\,{\textstyle\frac{4(d-3)}{d-2}}
\frac{b_1 B_0 - b_0 B_1}{b_0 b_2 - b_1^2}
h^\mu_\rho h^\nu_\sigma \, {\ocT}_{\mu\nu}\left( h_*^* \cM \right)
+{\textstyle\frac{4(d-3)}{d-2}}
\frac{b_1 B_1 - b_2 B_0}{b_0 b_2 - b_1^2}
h^\mu_\rho h^\nu_\sigma \, {\ocT}_{\mu\nu}\left( \cM \right)
\bigg\}
=0\, 
.
%\hspace{12pt}&&
\label{Thh_EEN2}
\end{eqnarray}
Performing the summations over $N$ in definitions (\ref{bm}) and (\ref{Bm}),
we obtain the following expressions for the parameters 
$b_m$ and $B_m$:
\begin{eqnarray}
b_0
=
&&\hspace{-18pt}
\alpha_1+2\,\alpha_2\,\cT\left(\cM\right)
,
\label{b0N2}
\\[2pt]
b_1
=
&&\hspace{-18pt}
\alpha_1(d-1)+2\,\alpha_2\,\cT\left(h_*^*\cM\right)
,
\label{b1N2}
\\[2pt]
b_2
=
&&\hspace{-18pt}
\alpha_1(d-2)(d-1)+2\,\alpha_2\,\cT\left((h_*^*)^2\cM\right)
,
\label{b2N2}
\\[2pt]
B_0
=
&&\hspace{-18pt}
-V+V'
+\alpha_1\,
\cT\left(\cM\right)
+\alpha_2
\left[
\,\cT\left(\cM^2\right)+{\textstyle\frac{2}{d-3}}\,\cT\left(\cP\cM\right)
-8\,\cT\left(\cN\right)
\,\right]
,\hspace{24pt}
\label{B0N2}
\\[2pt]
B_1
=
&&\hspace{-18pt}
-(d-1)V
+\alpha_1\,
\cT\left(h_*^*\cM\right)
+\alpha_2
\left[\,
\cT\left(h_*^*\cM^2\right)+{\textstyle\frac{2}{d-3}}\,\cT\left(h_*^*\cP\cM\right)
-8\,\cT\left(h_*^*\cN\right)\,
\right]
.
\label{B1N2}
\end{eqnarray}
Various generalized traces present in the above formulae 
should be replaced with the following explicit expressions:
\begin{eqnarray}
\cT\left(\cM\right)
&\!\!=&\!\!
R-K^2+(KK)+2(DD)\phi+2K\Ln\phi-(D\phi)^2-(\Ln\phi)^2
\,,
\label{cTcM}
%\end{equation}
%
%
%\begin{eqnarray}
\\[5pt]
\cT\left(\cM^2\right)
&\!\!=&\!\!
\left[
R - K^2 + (KK) +2(DD)\phi + 2K\Ln\phi - (D\phi)^2 - (\Ln\phi)^2
\right]^2
\nn[2pt]
&&
-\,4\,\big[
R_{\mu\nu} - KK_{\mu\nu} + (KK)_{\mu\nu} + (DD)_{\mu\nu}\phi
+ K_{\mu\nu}\Ln\phi 
\big]
\nn[2pt]
&& \hspace{0.55cm}
\cdot\big[
R^{\mu\nu} - KK^{\mu\nu} + (KK)^{\mu\nu} + (DD)^{\mu\nu}\phi
+ K^{\mu\nu}\Ln\phi 
\big]
\nn[2pt]
&&
+\,\big[
R_{\mu\nu\rho\sigma} - K_{\mu\rho}K_{\nu\sigma} + K_{\mu\sigma}K_{\nu\rho}
\big]\big[
R^{\mu\nu\rho\sigma} - K^{\mu\rho}K^{\nu\sigma} + K^{\mu\sigma}K^{\nu\rho}
\big]\,,
\label{cTcMcM}
\\[5pt]
\cT\left(\cP\cM\right)
&\!\!=&\!\!
{\textstyle\frac{2}{d-1}}\left[R-K^2+(KK)\right]
\left[R-K^2+(KK)+(DD)\phi+K\Ln\phi\right]
\nn[2pt]
&&
-\,2\,\big[
R_{\mu\nu}-KK_{\mu\nu}+(KK)_{\mu\nu}+(DD)_{\mu\nu}\phi+K_{\mu\nu}\Ln\phi
\big]
\nn[2pt]
&& \hspace{0.55cm}
\cdot
\big[
R^{\mu\nu}-KK^{\mu\nu}+(KK)^{\mu\nu}+(d-2)E^{\mu\nu}
\big]\, ,
\label{cTcPcM}
\\[5pt]
\cT\left(\cN\right)
&\!\!=&\!\!
\left[
D_{\mu}K-D_{\rho}K^{\rho}_{\mu}+K_{\mu}^{\rho}D_{\rho}\phi-D_{\mu}\Ln\phi
\right]
\big[
D^{\mu}K-D^{\sigma}K_{\sigma}^{\mu}+K^{\mu}_{\sigma}D^{\sigma}\phi-D^{\mu}\Ln\phi
\big]
\nn[2pt]
&&
-\,(D_{\mu}K_{\rho}^{\sigma})(D^{\mu}K^{\rho}_{\sigma})
+(D_{\mu}K_{\rho}^{\sigma})(D^{\rho}K^{\mu}_{\sigma})\,,
\label{cTcN}
\\[5pt]
\cT\left(h_*^*\,\cM\right)
&\!\!=&\!\!
(d-3)\big[R-K^2+(KK)\big]+2(d-2)\big[(DD)\phi+K\Ln\phi\big]
\nn[2pt]
&&
-\,(d-1)\big[(D\phi)^2+(\Ln\phi)^2\big]\,,
\label{cThcM}
%\end{eqnarray}
%
%
%\begin{eqnarray}
\\[5pt]
\cT\left(h_*^*\,\cM^2\right)
&\!\!=&\!\!
(d-5)\left[R - K^2 + (KK)\right]^2
+2(d-4)\left[R - K^2 + (KK)\right]
\big[(DD)\phi + K\Ln\phi\big]
\nn[2pt]
&&
+\,(d-3)\Big(
4\big[(DD)\phi + K\Ln\phi\big]^2
-\left[R - K^2 + (KK)\right]\left[(D\phi)^2 + (\Ln\phi)^2\right]
\Big)
\nn[2pt]
&&
-\,4(d-2)\big[(DD)\phi + K\Ln\phi\big]\big[(D\phi)^2 + (\Ln\phi)^2\big]
+(d-1)\left[(D\phi)^2 + (\Ln\phi)^2\right]^2
\nn[2pt]
&&
-\,4(d-5)\big[R_{\mu\nu} - KK_{\mu\nu} + (KK)_{\mu\nu}\big]
\big[R^{\mu\nu} - KK^{\mu\nu} + (KK)^{\mu\nu}\big]
\nn[2pt]
&&
-\,8(d-4)\big[R_{\mu\nu} - KK_{\mu\nu} + (KK)_{\mu\nu}\big]
\big[(DD)^{\mu\nu}\phi+ K^{\mu\nu}\Ln\phi \big]
\nn[2pt]
&&
-\,4(d-3)\big[(DD)_{\mu\nu}\phi+ K_{\mu\nu}\Ln\phi \big]
\big[(DD)^{\mu\nu}\phi+ K^{\mu\nu}\Ln\phi \big]
\nn[2pt]
&&
+\,(d-5)\big[
R_{\mu\nu\rho\sigma} - K_{\mu\rho}K_{\nu\sigma} + K_{\mu\sigma}K_{\nu\rho}
\big]\big[
R^{\mu\nu\rho\sigma} - K^{\mu\rho}K^{\nu\sigma} + K^{\mu\sigma}K^{\nu\rho}
\big]\,,
\label{cThcMcM}
%\end{eqnarray}
%
%
%\begin{eqnarray}
\\[5pt]
\cT\left(h_*^*\,\cP\cM\right)
&\!\!=&\!\!
%&&\hspace{-16pt}
{\textstyle\frac{2(d-4)}{d-1}}\left[R-K^2+(KK)\right]^2
+{\textstyle\frac{2(d-3)}{d-1}}\left[R-K^2+(KK)\right]
\big[(DD)\phi+K\Ln\phi\big]
\nn[2pt]
&&
-\,2\,\Big((d-4)\big[R_{\mu\nu}-KK_{\mu\nu}+(KK)_{\mu\nu}\big]
+(d-3)\big[(DD)_{\mu\nu}\phi+K_{\mu\nu}\Ln\phi\big]
\Big)
\nn[2pt]
&&\hspace{0.55cm}
\cdot
\big[
R^{\mu\nu}-KK^{\mu\nu}+(KK)^{\mu\nu}+(d-2)E^{\mu\nu}
\big]\,,
\label{cThcPcM}
\\[5pt]
\cT\left(h_*^*\,\cN\right)
&\!\!=&\!\!
(d-4)\Big(\left[D_{\mu}K-D_{\rho}K^{\rho}_{\mu}\right]
\big[D^{\mu}K-D^{\sigma}K_{\sigma}^{\mu}\big]
-(D_{\mu}K_{\rho}^{\sigma})(D^{\mu}K^{\rho}_{\sigma})
+(D_{\mu}K_{\rho}^{\sigma})(D^{\rho}K^{\mu}_{\sigma})
\Big)
\nn[2pt]
&&
+\,2\, (d-3)K_{\mu}^{\nu}\big[
(D_{\nu}K)(D^{\mu}\phi) - (D_{\rho}K^{\rho}_{\nu})(D^{\mu}\phi)
\big]
\nn[2pt]
&&
+\,(d-2)\left[K_{\mu}^{\rho}D_{\rho}\phi-D_{\mu}\Ln\phi\right]
\big[K^{\mu}_{\sigma}D^{\sigma}\phi-D^{\mu}\Ln\phi\big]\,,
\end{eqnarray} 
\begin{eqnarray}
%\\[5pt]
\cT\left((h_*^*)^2\cM\right)
&\!\!=&\!\!
(d-4)(d-3)\big[R-K^2+(KK)\big]+2(d-3)(d-2)\big[(DD)\phi+K\Ln\phi\big]
\nn[2pt]
&&
-\,(d-2)(d-1)\big[(D\phi)^2+(\Ln\phi)^2\big]\,,
%\end{eqnarray} 
%
%
%\begin{eqnarray}
\\[5pt]
h^\mu_\rho h^\nu_\sigma \, {\ocT}_{\mu\nu}\left( \cM \right)
&\!\!=&\!\!
h_{\rho\sigma}\,\cT\left(\cM\right)
-2  \big[ R_{\rho\sigma}-KK_{\rho\sigma}+(KK)_{\rho\sigma}
+(DD)_{\rho\sigma}\phi+K_{\rho\sigma}\Ln\phi\big] 
\,,
\\[5pt]
h^\mu_\rho h^\nu_\sigma \, {\ocT}_{\mu\nu}\left( \cP \right)
&\!\!=&\!\!
{\textstyle\frac{1}{d-1}}h_{\rho\sigma}\big[R-K^2+(KK)\big]
-\big[R_{\rho\sigma}-KK_{\rho\sigma}+(KK)_{\rho\sigma}\big]
\,,
\\[5pt]
h^\mu_\rho h^\nu_\sigma \, {\ocT}_{\mu\nu}\left(\cM^2\right)
&\!\!=&\!\!
h_{\rho\sigma}\,\cT\left(\cM^2\right)
-4 \big[R_{\rho\sigma}-KK_{\rho\sigma}+(KK)_{\rho\sigma} 
+(DD)_{\rho\sigma}\phi+K_{\rho\sigma}\Ln\phi\big] \,
\cT\left(\cM\right)
\nn[2pt] 
&& \hspace{-50pt}
+\,8\,\big[
R_{\rho\mu} - KK_{\rho\mu} + (KK)_{\rho\mu}  +  (DD)_{\rho\mu}\phi
+ K_{\rho\mu}\Ln\phi \big] 
\nn[2pt]
&& 
\cdot
\big[
{R^{\mu}}_{\sigma} - K{K^{\mu}}_{\sigma} + {(KK)^{\mu}}_{\sigma} 
+  {(DD)^{\mu}}_{\sigma}\phi + {K^{\mu}}_{\sigma}\Ln\phi \big]
\nn[2pt]
&& \hspace{-50pt}
+\,8\,\big[
R_{\rho\sigma\mu\nu} - K_{\rho\mu}K_{\sigma\nu} + K_{\rho\nu}K_{\sigma\mu}
\big]
\big[
R^{\mu\nu} - KK^{\mu\nu} + (KK)^{\mu\nu}  +  (DD)^{\mu\nu}\phi
+ K^{\mu\nu}\Ln\phi 
\big] 
\nn[2pt]
&& \hspace{-50pt}
-\,4\,\big[
R_{\rho\mu\nu\lambda} - K_{\rho\nu}K_{\mu\lambda} + K_{\rho\lambda}K_{\mu\nu}
\big]
\big[
{R_{\sigma}}^{\mu\nu\lambda} - {K_{\sigma}}^{\nu}K^{\mu\lambda} 
+ {K_{\sigma}}^{\lambda}K^{\mu\nu}
\big]\,,
\\[5pt]
h^\mu_\rho h^\nu_\sigma \, {\ocT}_{\mu\nu}\left(\cN\right)
&\!\!=&\!\!
h_{\rho\sigma}\,\cT\left(\cN\right)
+2(D_{\mu}K^{\nu}_{\rho})(D^{\mu}K_{\nu\sigma})
\nn[2pt]
&& \hspace{-50pt}
-\,2\,\left[D_{\mu}K-D_{\nu}K^{\nu}_{\mu}\right]
\big[D^{\mu}K_{\rho\sigma}-D_{\rho}K^{\mu}_{\sigma}\big]
+\left[D_{\rho}K_{\mu}^{\nu}-D_{\mu}K_{\rho}^{\nu}\right]
\big[D_{\sigma}K^{\mu}_{\nu}-D_{\nu}K^{\mu}_{\sigma}\big]
\nn[2pt]
&& \hspace{-50pt}
- \big[
D_{\rho}K-D_{\mu}K^{\mu}_{\rho}+K_{\rho}^{\mu}D_{\mu}\phi
-D_{\rho}\Ln\phi
\big]
\big[
D_{\sigma}K-D_{\nu}K^{\nu}_{\sigma} + K_{\sigma}^{\nu}D_{\nu}\phi
-D_{\sigma}\Ln\phi\big]
\,,\hspace{24pt}
\label{Tmunu_cN}
%\eea
%
%
%\bea
\\[5pt]
h^\mu_\rho h^\nu_\sigma \, {\ocT}_{\mu\nu}\left(h_*^*\cM\right)
&\!\!=&\!\!
h_{\rho\sigma}
\Big(
(d-4)\big[R-K^2+(KK)\big]+2(d-3)\big[(DD)\phi+K\Ln\phi\big]
\nn[2pt]
&& \hspace{0.9cm}
-\,(d-2)\big[(D\phi)^2+(\Ln\phi)^2\big]\Big)
\nn[2pt]
&& \hspace{-50pt}
-\,2\,(d-3)\big[R_{\rho\sigma}-KK_{\rho\sigma}+(KK)_{\rho\sigma} \big]
-2(d-2)\big[(DD)_{\rho\sigma}\phi+K_{\rho\sigma}\Ln\phi\big]
\,,\hspace{12pt}
\label{TmunuhcM}
\\[5pt]
h^\mu_\rho h^\nu_\sigma \, {\ocT}_{\mu\nu}\left(\cP\cM\right)
&\!\!=&\!\!
h_{\rho\sigma}\, \cT\left(\cP\cM\right)
-{\textstyle\frac{1}{d-1}}
h^\mu_\rho h^\nu_\sigma \, {\ocT}_{\mu\nu}\left(h^*_*\cM\right)
-\big[R_{\rho\sigma}-KK_{\rho\sigma}+(KK)_{\rho\sigma}+(d-2)E_{\rho\sigma}\big]
\nn[2pt]
&&\hspace{-50pt}
-\,2\,\big[R-K^2+(KK)\big]
\big[R_{\rho\sigma} - KK_{\rho\sigma} + (KK)_{\rho\sigma}
+ (DD)_{\rho\sigma}\phi + K_{\rho\sigma}\Ln\phi\big]
\nn[2pt]
&&\hspace{-50pt}
+\,2\,\big[R_{\rho\mu}-KK_{\rho\mu}+(KK)_{\rho\mu}+(d-2)E_{\rho\mu}\big]
\big[R_\sigma^\mu - KK_\sigma^\mu + (KK)_\sigma^\mu
+ (DD)_\sigma^\mu\phi + K_\sigma^\mu\Ln\phi\big]
\nn[2pt]
&&\hspace{-50pt}
+\,2\,\big[R_\sigma^\mu-KK_\sigma^\mu+(KK)_\sigma^\mu+(d-2)E_\sigma^\mu\big]
\big[R_{\rho\mu} - KK_{\rho\mu} + (KK)_{\rho\mu}
+ (DD)_{\rho\mu} + K_{\rho\mu}\Ln\phi\big]
\nn[2pt]
&&\hspace{-50pt}
+\,2\,\big[R_{\rho\mu\sigma\nu} - K_{\rho\mu}K_{\sigma\nu}
+K_{\rho\nu}K_{\sigma\mu}\big]
\big[R^{\mu\nu} - KK^{\mu\nu} + (KK)^{\mu\nu} + (d-2)E^{\mu\nu} \big]
\,.
\label{TmunucPcM}
\end{eqnarray}
By substituting eqs.\ (\ref{b0N2})-(\ref{TmunucPcM}) 
into (\ref{Thh_EEN2}) and employing the solutions of the junction 
conditions (\ref{tensorJCN2}) and (\ref{scalarJCN2}),
the effective brane equation of motion in 
the Einstein-like form can be obtained. However, the result is quite
intricate and will not be given here. 
Moreover, the formulae collected in this appendix should be compared 
with the relatively simple eq.\ (\ref{TnnEoMN2}).
This shows how big price, in terms of complication, has to be
paid in order to derive the effective brane equation of motion 
in the Einstein-like form, instead of confining to eq.\ (\ref{TnnEoMN2}).

Similarly to the $N_{\rm max}=1$ case, the whole bulk dependence 
of the brane dynamics is encoded in the projected Weyl 
tensor $E_{\mu\nu}$. However, this dependence is quite complicated
for $N_{\rm max}=2$. Specifically, $E_{\mu\nu}$ enters 
the brane Einstein-like equation of motion (\ref{Thh_EEN2}) through 
the generalized trace 
$h^\mu_\rho h^\nu_\sigma \, {\ocT}_{\mu\nu}\left( \cP \cM \right)$,
appearing in eq.\ (\ref{Thh_EEN2}) and given by (\ref{TmunucPcM}),
as well as through the generalized traces $\cT(\cP\cM)$ and $\cT(h_*^*\cP\cM)$,
present in the definitions of $B_0$ and $B_1$ and given by 
(\ref{cTcPcM}) and (\ref{cThcPcM}), respectively. 
Although the projected Weyl tensor $E_{\mu\nu}$
enters these formulae only linearly, it is involved in intricate contractions 
with other tensors. Furthermore, these tensors contain the extrinsic 
curvature $K_{\mu\nu}$, i.e.\ a complicated solution 
of the junction conditions (\ref{tensorJCN2}) and (\ref{scalarJCN2}) 
- which in general cannot be solved explicitly. 

As for $N_{\rm max}>2$, it is in principle possible to write down 
the Einstein-like effective brane equation of motion explicitly, 
but they become practically intractable.

%%%%%%%%%%%%%%%%%%%%%%%%%%%%%%%%%%%%%%%%%%%%%%%%%%

\end{document}